\documentclass[]{article}
\usepackage[utf8]{inputenc}
\usepackage[T1]{fontenc}
\usepackage{amssymb}
\usepackage{graphicx}
\usepackage{multirow}
\usepackage{lscape}
\usepackage{rotating}
\usepackage{enumitem}
\usepackage{float}
\usepackage{booktabs,tabularx}

\title{Cytology Image Analysis Techniques Towards Automation: Systematically Revisited}
\author{Shyamali Mitra\thanks{Department of Instrumentation and Electronics Engineering, Jadavpur University}, Nibaran Das\thanks{Department of Computer Science \& Engineering, Jadavpur University}, , Soumyajyoti Dey\thanks{Department of Computer Science \& Engineering, Jadavpur University}\\ Sukanta Chakrabarty\thanks{Theism Medical Diagnostics Centre}, Mita Nasipuri\thanks{Department of Computer Science \& Engineering, Jadavpur University}, Mrinal Kanti Naskar\thanks{Department of Electronics \& Telecommunication Engineering, Jadavpur University}}

\begin{document}

\maketitle

\begin{abstract}
	
	Cytology is the branch of pathology which deals with the microscopic examination of cells for diagnosis of carcinoma or inflammatory conditions.
	Automation in cytology started in the early 1950s with the aim to reduce manual efforts in diagnosis of cancer. The inflush of intelligent technological units with high computational power and improved  specimen collection techniques helped to achieve its technological heights. In the present survey, we focus on such image  processing techniques which put steps forward towards the automation of cytology. We take a short tour to 17  types of cytology and explore various segmentation and/or classification techniques which evolved during last three decades boosting the concept of automation in cytology. It is observed, that most of the works are aligned towards three types of cytology: Cervical, Breast and Lung, which are discussed elaborately in this paper. The user-end systems developed during that period are summarized to comprehend the overall growth in the respective domains. To be precise, we discuss the diversity of the state-of-the-art methodologies, their challenges to provide prolific and competent future research directions inbringing the cytology-based commercial systems into the mainstream.

\end{abstract}

\section{INTRODUCTION}
\begin{figure}[h]
	\centering
	\includegraphics[width=0.5\linewidth]{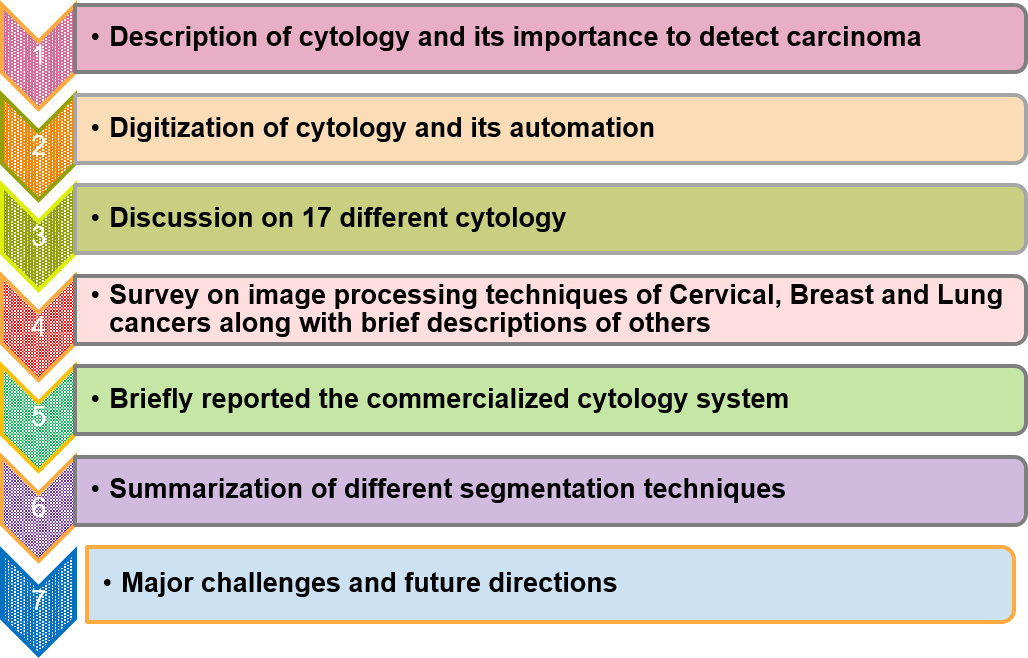}
	\caption{{Highlights of  the present Survey }}
	\label{fig:1}
\end{figure}

%\label{S:1}
%\label{sec1}
Cancer, as of now, has become a soaring concern in individuals. Living organisms are composed
of large number of cells. Normal cells have specific lifetime in which they reproduce and divide to
produce new cells, replacing the old ones. Before the old cells worn out it passes its genetic
information to the new ones. Thus, cell divisions are initiated by a control signal generated from a
specific protein called cyclins. But, cancer cells divide in a proliferative manner thereby violating
coordination among  cells. Though cancer causing genes are inherited genetically, various
external factors like indulging
in smoking and drinking, exposure to heavy metals, radiation, usage of plastics etc. \cite{Danaei2005CausesFactors} are also responsible for this. Cancer cells
have several distinguishing characteristics from benign cells that are extensively discussed in different
types of cytology. {More specifically, the study of tissues and cells using a microscope to detect malignancies or other inflammatory conditions is broadly categorized under cytology or cell biology.} There are several challenges {in  analyzing cytology images} which make the overall diagnosis
process difficult even by a trained and experienced cytotechnologists. {The challenge percolates from  nuclear shape, size, density to diverse nature of cytoplasm \cite{cibas2013cytology}}. 
There are different automatic or semi-automatic systems \cite{Street2000Xcyt:Cancer,Yang2004DetectionCancer.,Carpenter2006,Kemp2007DetectionCytometry,Wilbur2011DigitalFuture,Ghosh2013ValueCases,George2014RemoteImages,Litjens2017AAnalysis, Zhang2017DeepPap:Classification} to aid cytotechnologists in the diagnosis procedure  .
The existing survey papers \cite{Masood2011DiagnosticCare, Saha2016Computer-aidedReview,WILLIAM201815} address the segmentation and classification problem for a single domain in cytology. Also it is observed that meagre attention is given to cytology compared to histopathology {despite its potential to classify a malignancy cell in the least invasive manner}. Therefore, it is very difficult to gather an overall knowledge about the entire domain
of cytology, related to the challenges and the limitations. This provoked the idea of excavating cytology based research works to give better pace  in this domain. In this paper, we
attempt to produce a comprehensive list of 17 different types of cytology based on their sites of origin.
We also discuss a large pool of segmentation and classification algorithms with associated challenges in meeting
higher performance requirements, outline techniques to address common bottlenecks, correlation
between them and provide meaningful insights on the techniques for future developments for
automation in the diagnosis process. The research works over past 30 years in these domains of cytology are systematically and extensively reviewed so that one can have a deeper insight into the evolution of
methodologies.
%The purpose of the study, again, is not intended to provide heuristic study of image segmentation and classification algorithms, but a correlated study on the underlying principles specific to image properties. 
The highlights of the present survey illustrated is given in Fig.\ref{fig:1}. 
%The organization of the paper is crafted according to the steps required in Computer-aided
%diagnosis (CAD) system as shown in Fig.\ref{fig:2}. 
In section 2, we provide a brief introduction to
cytology. In section 3, we discuss automation in cytology and its
underlying principles of image analysis. In section 4, we discuss briefly about 17 different types of
cytology and undergo a concise and comprehensive survey on three different domains of cytology viz.
cervical, breast, and lung, where extensive research works on the
advanced techniques for segmentation and classification problems are carried out.% Rest of the 14 
Other types of cytology
%where a very few research works are encountered
are grouped under miscellaneous section. In section 5, we mentioned the state-of-the-art systems developed to comprehend overall progress in the respective domains. A section  based on authors' view is presented   to discuss concisely about the comprehensive progress of cytology image analysis techniques.  

\section{CYTOLOGY: A BRIEF INTRODUCTION}

%\subsection{Subsection One}

The  body of an organism is comprised of millions and trillion of cells. Each cell possesses a cytoplasm and a nucleus. Cytoplasm or the cell body acts as an envelope to the nucleus containing
chromosomes, the genetic material that undergoes mutation under certain changes in the form of
diseases. The changes are reflected in the morphology of nucleus and cytoplasm. Thus, the micro
examination of cells, can unfold relevant information on any morphological alteration in
response to a particular disease. {These cells, which are important predictors of pre-malignant and
	malignant lesions, can be sampled and examined under the microscope to diagnose different medical
	conditions with the exception of a few. Cytology test is an easy option for the patients because of its painless procedure to detect
	the disease and also an attractive option for the doctors because it is easier to treat the disease at the
	nascent stage. In certain tests, \cite{pannek2017clinical} cytology is sensitive to potentially high grade tumors and  cytology when combined with other modalities of testing can increase the sensitivity of detection to a greater extent.

	In benign cells, the cellular materials are  generally well-defined,  exhibiting uniform chromatin distribution within the nucleus  and a prominent cytoplasm. The  nucleus boundary is usually regular and nearly elliptical in shape. Malignant cells, on the other hand, possess irregular nuclear boundary. Scanty cytoplasm and multiple nuclei with unusual sizes can be observed in a single cell. For classifying a specimen into benign and malignant, these characteristics are usually taken under consideration. 
	
	Histopathology gives information of tissue for further immunohistological and molecular analysis needed for targeted cancer chemotherapy.  In breast cancer, Fine Needle Aspiratuin Cytology (FNAC) is pretty good in diagnosing carcinoma and benign fibroadenomas, but histopathology needs some borderline and stromal lesions. In thyroid cell carcinoma, FNAC can diagnose almost all types of goitre, thyroiditis, papillary carcinoma, medullary carcinoma etc. Histopahology, on the other hand, is necessary to differentiate  follicular adenoma and carcinoma. In lung cancer, FNAC is best suited to detect infections and primary diagnosis of cancers. Histopathology of lung  may be harmful in some infective conditions like tuberculosis etc.
	%Lung histopahology is  much more risky than FNA. In case of lymph nodes, reactive nodes are well differentiated by TB-FNA (Transbronchial FNA). But for lymphomas, histopathology  followed by immunohistochemistry is very essential to obtain a good sensitivity and accuracy from diagnostic point of view.}
	
	Two types of cytology tests are undertaken to determine presence of carcinoma and if detected,
	figures out the extent of disease. A screening test may be performed at regular intervals to assure the presence of disease. It is helpful to detect the disease at
	at an infant stage when response to the treatment is faster and easier  so that  chances of
	fatality is reduced. It is generally recommended for those who are highly prone to this disease. A diagnostic test is
	done when signs or symptoms are more common. This test is carried out to check the presence of
	disease and if so, aimed at precise identification of the specific condition of the disease.
	
	Preceding to any screening or the diagnostic test, the specimen needs to be prepared. The preparation
	consists of three steps: i) Specimen collection, ii) Slide Preparation and iii) Fixation, which are described in detail in supplementary section. 
	\section{\textbf{{PRINCIPLES OF CYTOLOGY IMAGE ANALYSIS}}}
	
	After a suitable slide is prepared, a screening or a diagnostic test can be performed for the identification
	of the specimen. It can be done manually or through automated CAD based system. A manual screening
	procedure directly involves an expert cytotechnologist to review/examine the slides. 
	%After proper examination of the slide under microscope, the final verdict regarding the nature of the specimen is produced by the doctors/medical experts. 
	Whereas, in automatic screening system, there are few steps
	that are handled by a computer using some image processing protocols. For  automated systems,
	result is predicted by a locally installed or any remotely installed device without human counterpart. A comparative
	pictorial representation of manual and automatic screening system is given in Fig. \ref{fig:4}.
	\begin{figure*}[h]
		\centering
		\includegraphics[width=0.7\linewidth]{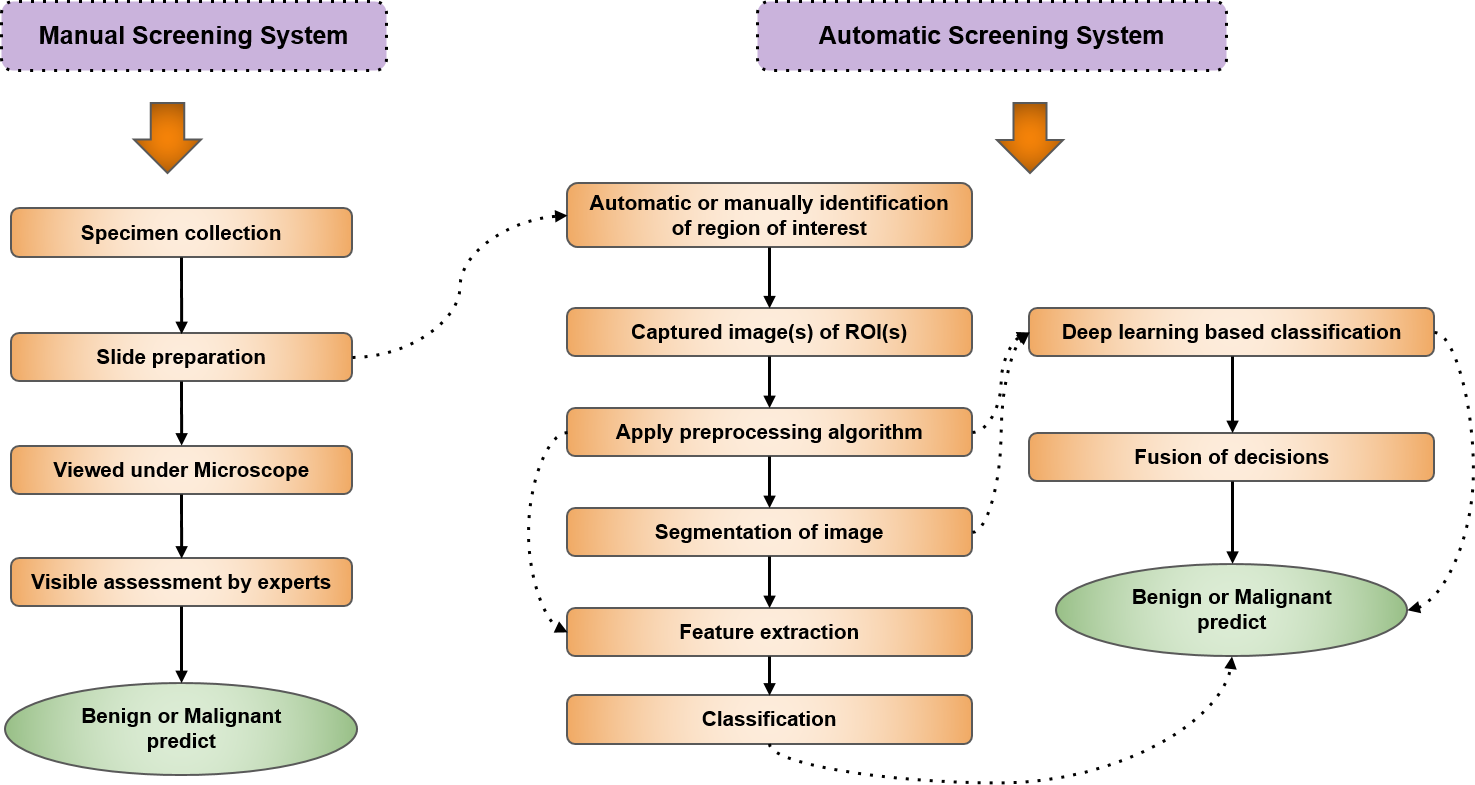}
		\caption{{Pictorial representation of manual and automatic malignant screening system}}
		\label{fig:4}
	\end{figure*}
	{It is  worthy to mention here that there are certain factors that hinder the proper interpretation of slide leading to an erroneous result. Such factors include:}
	a) {Sample inadequacy or fewer number of malignant cells},
	b) {Poor preparation of slide. Slide may contain excessive blood, mucus, inflammatory cells, background necrosis or other foreign particles and obscuring materials making overall appearance can unclear and cloudy},
	c){Poorly smeared or stained},
	d) {Poor preservation of slides},
	e) {Lack of experience of pathologists to collect the specimen from the exact location of the tumor},
	\\ {All these factors give rise to  what is known as sampling error or otherwise stated purging of these errors is a prerequisite for exact rendition of a sample.}
	
	%clearpage
	\subsection{\textbf{The automation as a growing concept in cytology}}
	
	Since long back, image analysis using visual interpretation has been  the primary essence in cytological image
	analysis. The result is assessed based on the examination of cells under microscope by skilled
	practitioners which is a dearth as of now. 
	%So, reduction in human intervention with the help of automated devices has become necessary in cytology laboratory in the current scenario. 
	Therefore, the  objective
	behind the soaring trend of computer assisted diagnostic system is to reduce the time taken to perform
	the tests  and complete the process of report generation. There are  other factors like reducing the
	workload of cytotechnologists,  human induced errors, allowing batch processing etc. which
	bolstered the motivations further to put steps forward towards automation. The aim to develop an
	automated screening system can be summarized into two categories:
	
	\begin{itemize}
		\item   It can act as pre-screening system to differentiate a benign and a malignant specimen.
		To eradicate false negative cases, such systems must exhibit high false positive cases, so
		that, no single abnormal specimen  demands further investigation by doctors are
		disregarded. Thus, in another sense, it can act as an adjunct to the cytotechnologists by
		eliminating the need of evaluating normal specimens, which saves time and energy as well.
		\item
		%Another way to implement an automated system is 
		To run a parallel system  to the
		present manual screening system. Thus, the diagnosis process can be bi-directed to produce
		less chances of errors at the specimen level after fusion.
	\end{itemize}

The immediate impact of automation in cytology, that is realized by the entire community of doctors, is
that it allows batch processing, eliminating the manual tiring and time consuming process. During early
1950$'$s, the idea to develop ‘Cytoanalyzer’ \cite{Diacumakos1962ExfoliatedCytoanalyzer}, an automated scanning device, was the initial
breakthrough to the concept of manual examination of slides. This could distinguish normal cells from abnormal
cells using the information of nuclear size and optical density. However, the system suffered from two
major disadvantages. The process was slow due to low processing power of the computers at that time. Apart from that, overlapping of cells and air drying factors of conventional
pap smears complicated the identification procedure further with higher false positive rates. Atrophic changes in conventional smears of older patients added further hurdles to classify the images
correctly by using automated analysis. So, there were several issues related to conventional pap smear
slides which hindered the correct analysis by computer aided systems. To mitigate the issues, Zahniser et al.
{\cite{Zahniser1996AutomatedLaboratory}} % 
came up with the idea of liquid cell suspension to make “a better pap test” of the cervical
sample. They used Feulgen stain rather than Papanicoloau stain. The software developed in conjunction
with liquid based specimens gave better results with a dramatic decrease in false negative rates to 1\%
and false positive rates to 10\%.

Another angle to successful implementation of automated screening system was to incorporate more shape and texture based
features of nuclei and cytoplasm to achieve more accuracy and robustness.
This idea was given a shape by Toshiba with the name of CYBEST \cite{Watanabe1974AnCYBEST}. It was a prototype developed
in 1972 with further development in a new desk size design in 1981. AutoPap 300 and SurePath
were two automated screening systems that were marketed successfully.
\begin{figure*}[h]
	\centering
	\includegraphics[width=0.85\linewidth]{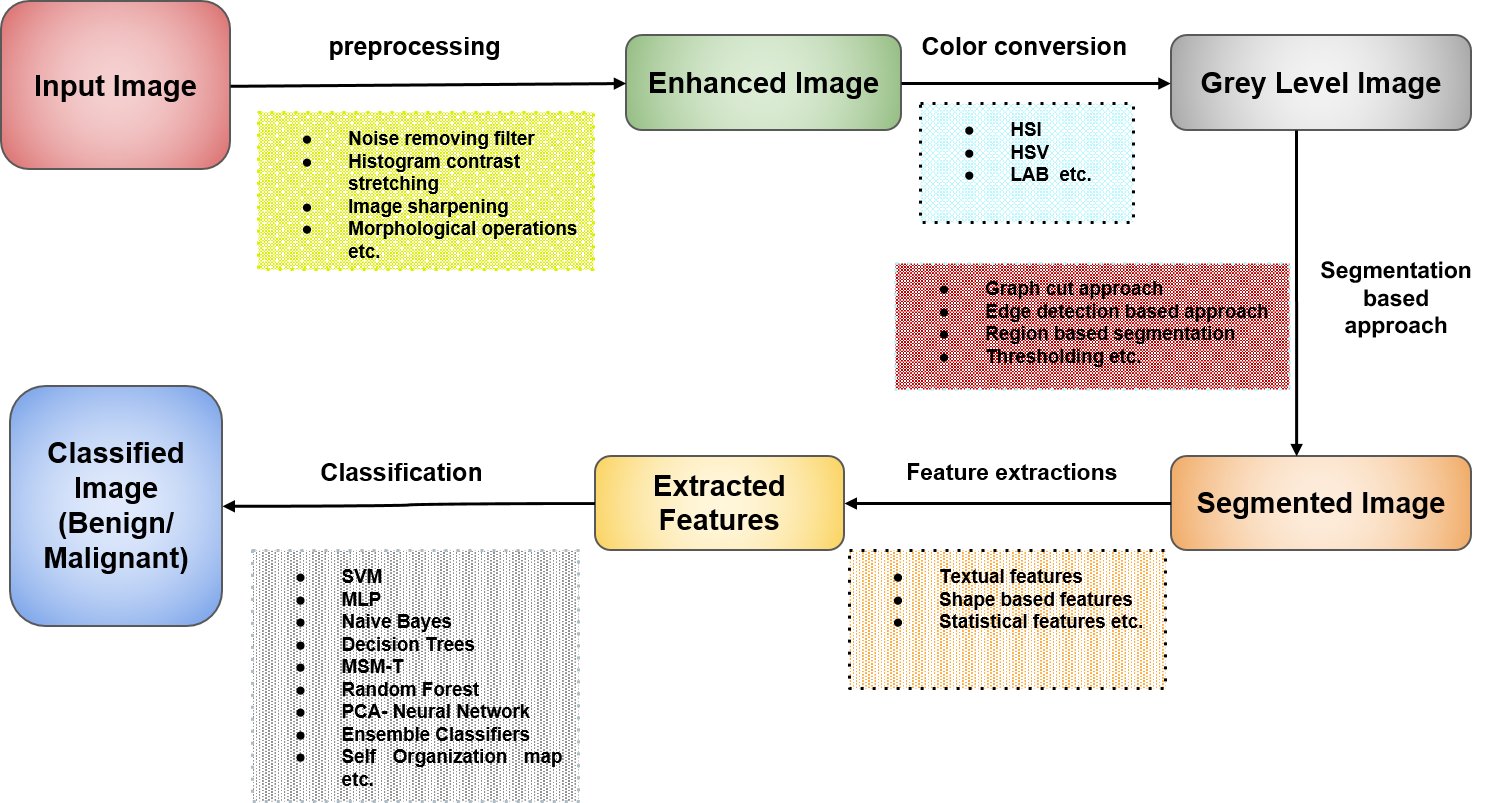}
	\caption{{Image analysis steps of a typical segmentation based image processing system where the intermediate stages
			denote the processing techniques}}
	\label{fig:5}
\end{figure*} 
Despite limited success achieved so far with the existing systems, there were constant efforts to upgrade
and design automated screening systems on a large scale. There were a handful of designs developed
during 1980's like {CERVISCAN \cite{Tucker1976Cerviscan:Prescreening}, LEYTAS \cite{Al1979DetectionSystem}, FAZYTAN \cite{Erhardt1980FAZYTAN:Processing.}, BioPEPR \cite{Oud1981TheSystem.} etc.}

In cytology laboratories there are three major sections  where automations are
extremely pertinent and devices are currently accessible as follows:

\begin{itemize}
	\item Specimen preparation devices:  Two FDA(Food and Drug Administration) approved  automated systems 
	namely, ThinPrep Processor and
	the AutoCyte Prep serve the purpose popularly. ThinPrep processor 5000 uses thin prep technology, for cell dispersion,
	collection and transfer, which can process approximately 35 slides/hour. 
	The preparation of monolayer slides in AutoCyte prep is done using liquid based preparations.
	\item Manual screening adjunctive devices: These devices are used either independently or in
	conjunction with human by focusing the portion of slides having abnormalities. Thus,
	examining the whole slide is not required. Some of the computerized
	FDA approved microscopes such as CompuCyte’s "M Pathfinder", Accumed International’s "AcCell 2000" 
	are widely used to mark abnormal cells.
	%\clearpage
	
	\item Automated screening devices: To automate the process the interpretation of cytology smears,
	Auto pap screener system, an existing system is designed to present a portion of slide that is
	adequate for analysis by cytotechnologist, thus trimming their workload to go through whole
	specimen.
\end{itemize}
\subsection{\textbf{Image analysis steps}}

The journey towards automation has lead researchers to enunciate new image
processing algorithms for better understanding and analysis of the images. During analysis of a cytology specimen, some
basic steps need to be followed. In manual screening system of cytology images, the steps include,
preparation of cytology specimen, feeding, fixing and adjusting the slide with required magnification in the microscope to identify the region of interest. After proper examination based on the nature
of specimen opinion is given by experts. Thus, each of these steps involve a human effort. In automatic
screening system, all the above steps required prior to image analysis can be automated or semi-automated, including image digitization, segmentation of the images, features extraction and finally providing the diagnostic information through judicious use of classifiers. Each of these steps {as mentioned in Fig. \ref{fig:5}}
gives rise to a number of challenging research problems that is studied in great deal over the last half
century since the achievements in the field took place. If the goal is to create an interactive system,
some of the difficulties can be left over human operator to solve, but when a fully automated system is
a goal, all the steps are needed to be addressed. In the next section, we focus  only  on automation techniques available in image analysis steps.

 \section{\textbf{DIFFERENT TYPES OF CYTOLOGY}}

There are different types of cytology based on the originating site of the human body. In  Table \ref{tab:new}, we
have summarized 17 such types of cytology along with the modalities are summarized. It has been observed that most of the automated approaches are devised for
screening of Cervical, Breast and Lung cancer.Therefore, those three cytology are  discussed in details. Each of the
work discussed in the next section throws light on how succeeding
works evolve to better and robust
algorithms. This will help the reader to get an overview about the domain very quickly. Some images of different cytology domains are shown in Fig.\ref{fig:im1},\ref{fig:im3},\ref{fig:im5},\ref{fig:7} for better understanding of the type of specimen.


% \usepackage{graphicx}

% \usepackage[normalem]{ulem}


\begin{table}[]

	\caption{Different types of cytology with their modalities and corresponding overall performances in detecting carcinoma}

	\label{tab:new}

	\resizebox{\textwidth}{!}{%

		\begin{tabular}{|c|c|c|c|l|l|l|l|}

			\hline

			\textbf{SL\#} & \textbf{Cytology} & \textbf{Used to detect} & \textbf{\begin{tabular}[c]{@{}c@{}}Modality of\\ specimen\\ collection\end{tabular}} & \multicolumn{1}{c|}{\textbf{\begin{tabular}[c]{@{}c@{}}Accuracy\\ (Acc)/Sensitivi\\ ty(Se)/Specific\\ ity(Sp)\end{tabular}}} & \multicolumn{1}{c|}{\textbf{\begin{tabular}[c]{@{}c@{}}Cytomorphology of\\ benign cells (General\\ characteristics irrespective\\ of sub types)\end{tabular}}} & \multicolumn{1}{c|}{\textbf{\begin{tabular}[c]{@{}c@{}}Cytomorphology of\\ Limitations of\\ malignant cells(General\\ characteristics irrespective\\ of sub types)\end{tabular}}} & \multicolumn{1}{c|}{\textbf{Limitations of cytology}} \\ \hline

			1 & Adrenal gland & To detect adrenal nodule. & Adrenal FNA & Acc : 96\%-98\% & \begin{tabular}[c]{@{}l@{}}a) Many naked nuclei.\\ b) Granular background.\\ c) Intact cells with fizzy cytoplasm\end{tabular} & \begin{tabular}[c]{@{}l@{}}a) Isolated cells with dense cytoplasm.\\ b) Pronounced nuclear atypia.\\ c) Mitoses.\\ d) Background necrosis.\end{tabular} & \begin{tabular}[c]{@{}l@{}}a) For nodules \textless{}3cm diagnostic\\ accuracy drops\end{tabular} \\ \hline

			2 & Breast & To detect breast tumors & \multicolumn{1}{l|}{\begin{tabular}[c]{@{}l@{}}a) Nipple discharge(ND).\\ b) Ductal lavage using FNA\end{tabular}} & \begin{tabular}[c]{@{}l@{}}ND :\\ Se :41\% - 60\%\\ FNA :\\ Se :65\% - 98\%\\ Sp :34\%-100\%\\ F.P :0\% -2\%\end{tabular} & \begin{tabular}[c]{@{}l@{}}Round to oval monomorphic\\ epithelial cell\\ and adherent dark nucleus.\\ (See Fig.7.i.a)\end{tabular} & \begin{tabular}[c]{@{}l@{}}nuclei with irregular boundaries.\\ b) Irregular spacing.\\ c) Background debris.\\ (See Fig.7.ii.b)\end{tabular} & \begin{tabular}[c]{@{}l@{}}a) Nipple discharge cytology is not\\ very effective.\\ b) It is performed to very few patients\\ who are asymptotic.\end{tabular} \\ \hline

			3 & Cerebrospinal fluid & To detect malignancy & \multicolumn{1}{l|}{\begin{tabular}[c]{@{}l@{}}Lumbar puncture by using\\ needle in intervertebral\\ space at L3to L4 or\\ L4 to L5.\end{tabular}} & \begin{tabular}[c]{@{}l@{}}Se :60\%\\ Sp : high\end{tabular} & \begin{tabular}[c]{@{}l@{}}a) Sparsely cellular\\ b) Round nuclei.\\ c) Uniform nuclear contours\end{tabular} & \begin{tabular}[c]{@{}l@{}}a) Irregular nuclear contours.\\ b) Prominent nucleolus\\ c)Abnormal chromatin pattern\end{tabular} & \begin{tabular}[c]{@{}l@{}}a) Additional processing is needed to\\ preserve the specimen\end{tabular} \\ \hline

			4 & Cervical cytology & To detect cervical cancer & \begin{tabular}[c]{@{}c@{}}Scrape or brush\\ cytology\end{tabular} & \begin{tabular}[c]{@{}l@{}}Se :47\%\\ Sp :95\%\end{tabular} & \begin{tabular}[c]{@{}l@{}}Large\\ polygonal\\ cellswith small round\\ nuclei and large amount\\ of eosinophelic\\ cytoplasm.\\ (See Fig.7.ii..a)\end{tabular} & \begin{tabular}[c]{@{}l@{}}Malignant melanoma :\\ a) Clusters of small cells.\\ b) Scant cytoplasm\\ c) Nuclear hyperchromasia\\ (See Fig.7.ii.b)\end{tabular} & \begin{tabular}[c]{@{}l@{}}a) False positive cases are more.\\ b) Less useful for cystic and fibrotic lesion\end{tabular} \\ \hline

			5 & Gastrointestinal cytology & \begin{tabular}[c]{@{}c@{}}To evaluate gastro-intestinal\\ tract lesions\end{tabular} & Brush cytology & \begin{tabular}[c]{@{}l@{}}Se :77\%-94\%\\ Sp :95\%\end{tabular} & \begin{tabular}[c]{@{}l@{}}a) Uniformly arranged\\ cellular structure.\\ b) Clear nuclear\\ boundary.\end{tabular} & \begin{tabular}[c]{@{}l@{}}a) Isolated cells and crowded groups,\\ b) Nuclear pleomorphism\\ c) Hyperchromasia.\\ d) irregular cell size and shape.\end{tabular} & \begin{tabular}[c]{@{}l@{}}a) Sampling error.\\ b) Accuracy compromised in deter-mining\\ Barrett’s esophagus.\\ c) High false positive cases.\end{tabular} \\ \hline

			6 & Kidney & To detect renal lesions & Renal FNA & Acc :73\% -94\% & \begin{tabular}[c]{@{}l@{}}a) Large dense globular\\ structures.\\ b) Prraorxei.mal tubular cells are rare\end{tabular} & \begin{tabular}[c]{@{}l@{}}a) Large cells.\\ b) Large nuclei,\\ c) Abunadant granular\\ cytoplasm\end{tabular} & \begin{tabular}[c]{@{}l@{}}a) FNAChas low sensitivity\\ for small masses(\textless{}5cm) and\\ complex cysts and\\ large masses \textgreater{}5cm\end{tabular} \\ \hline

			7 & Liver & \begin{tabular}[c]{@{}c@{}}To detect focal lesions\\ of liver.\end{tabular} & FNA & \begin{tabular}[c]{@{}l@{}}Se :71\%-91\%\\ Sp :87\% -100\%\\ Acc :90\%-94\%\end{tabular} & \begin{tabular}[c]{@{}l@{}}Hepatocytes are large,\\ polygonal, isolated cells\\ with prominent nucleolous\\ and binucleated.\\ They are centrally placed\\ and round to oval in shape.\\ (See Fig.4.ii.a)\end{tabular} & \begin{tabular}[c]{@{}l@{}}a) Isolated cells or cells in\\ nest. Increased nucleus to\\ cytoplasm ratio.\\ b) Large,round nucleus\\ with prominent nucleolus.\\ c) Often Large naked nuclei\\ present.\\ (See Fig.4.ii.b)\end{tabular} & \begin{tabular}[c]{@{}l@{}}a) Rare hemorrhage, pain,\\ bile peritonitis.\\ b) Also, cannot typify the tumors\end{tabular} \\ \hline

			8 & Lymph nodes & \begin{tabular}[c]{@{}c@{}}To confirm enlarged\\ lymph nodes\end{tabular} & FNA & \begin{tabular}[c]{@{}l@{}}Se : 80\%\\ Sp : 90\%\\ Acc : 93\%\end{tabular} & \begin{tabular}[c]{@{}l@{}}a) Dispersed isolated cell\\ pattern.\\ b) Presence of\\ lymphoglandular bodies.\end{tabular} & \begin{tabular}[c]{@{}l@{}}a) Polymorphus population.\\ b) Mast cells present.\\ (See Fig.6.ii)\end{tabular} & \begin{tabular}[c]{@{}l@{}}a) Sampling error.\\ b) Cell vascular\\ patterns are lost to some extent.\end{tabular} \\ \hline

			9 & Ovarian lesions & To detect cyst ovarian  masses & FNA & \begin{tabular}[c]{@{}l@{}}Se(for\\ borderline\\ analysis) :84\% -\\ 93\%\\ Se : 26 -40\%\end{tabular} & \begin{tabular}[c]{@{}l@{}}a) Sparsely to highly\\ cellular.\\ b) Cells possess\\ round nucleus with\\ coarsely granular\\ chromatin with one or\\ two small nucleoli.\end{tabular} & \begin{tabular}[c]{@{}l@{}}a) Clusters and isolated\\ cells.\\ b) Large\\ pleomorphic cells.\\ c) Round nuclei with\\ prominent nucleoli.\\ (See Fig.6.iii\end{tabular} & \begin{tabular}[c]{@{}l@{}}a) Higher false negative rates.\\ b) Lesser reliability\\ in distinguishing borderline\\ tumor and carcinoma.\\ c) Benign lesions not histolo-gically\\ categorized.\end{tabular} \\ \hline

			10 & Pancreas and biling tree & \begin{tabular}[c]{@{}c@{}}To indicate a pancreatic mass\\ and a duct stricture\end{tabular} & FNA and Duct Brushings & \begin{tabular}[c]{@{}l@{}}FNA :\\ Se : 86\% -98\%\\ Sp : 100\%\end{tabular} & \begin{tabular}[c]{@{}l@{}}a) Round or oval nucleus.\\ b) Evenly distributed\\ nucleus.\\ c) Finely\\ granulated chromatin.\\ d) Cytoplasm boundaries\\ well defined.\end{tabular} & \begin{tabular}[c]{@{}l@{}}a) Moderate to high cellularity.\\ b) Crowded sheets of disordered\\ negative cases. ductal cells.\\ c) Irregular nuclear contours.\\ d) Enlarged nucleolous.\\ e) Irregular\\ chromatin\\ distribution.\end{tabular} & \begin{tabular}[c]{@{}l@{}}a) Sampling error can lead\\ to false\end{tabular} \\ \hline

			11 & Peritoneal washing & \begin{tabular}[c]{@{}c@{}}To detect spread of cancer\\ in peritoneal surface\end{tabular} & Washing cytology & \begin{tabular}[c]{@{}l@{}}F.N :31\%-86\%.\\ F.P : \textless{}5\%\end{tabular} & \begin{tabular}[c]{@{}l@{}}a) Isolated mesothelial\\ cells in sheets, often folded.\\ b) Pale chromatin,\\ c) Small round or oval\\ nucleoli\end{tabular} & \begin{tabular}[c]{@{}l@{}}a) Cell\\ clusters\\ and\\ isolated cells.\\ b) Enlarged nuclei,\\ coarse chromatin.\\ c) Nuclear pleomorphism\\ d) Scant or abundant and\\ vacuolated cytoplasm.\end{tabular} & a) Sampling error \\ \hline

			12 & \begin{tabular}[c]{@{}c@{}}Pleural,Peri cardial and\\ Peritoneal fluid\end{tabular} & \begin{tabular}[c]{@{}c@{}}To assess the fluid in\\ diseased condition\end{tabular} & FNA & \begin{tabular}[c]{@{}l@{}}Sp : High\\ Se :71\%\\ F.P :1\%\\ F.N : High\end{tabular} & \begin{tabular}[c]{@{}l@{}}a) Numerous cells often dispersed.\\ b) Round nucleus with one nuc-leolus.\\ Dense cytoplasm\\ with clear boundary.\\ (See Fig.5.ii.a)\end{tabular} & \begin{tabular}[c]{@{}l@{}}a) Numerous\\ large clusters.\\ b) Presence of cell\\ block sections.\\ c)Second population\\ (See Fig.5.ii.b)\end{tabular} & \begin{tabular}[c]{@{}l@{}}a) False findings are high.\\ b) Immuno-cytochemistry\\ can act as companion to\\ improve diagnostic accuracy.\end{tabular} \\ \hline

			13 & Respiratory trac & \begin{tabular}[c]{@{}c@{}}To detect abnormalities\\ in respiratory tract.\end{tabular} & \begin{tabular}[c]{@{}c@{}}Sputum, Bronchial brushing,\\ Transbronchial FNA,\\ Trans esophageal FNA.\end{tabular} & \begin{tabular}[c]{@{}l@{}}Sputum :\\ Se :72\%\\ Sp :74\%\\ Transbronchial\\ FNA+bronchial\\ brushing :\\ Se :91\%\\ Sp :96\%-99\%\\ Acc :80\% -85\%\end{tabular} & \begin{tabular}[c]{@{}l@{}}a) Large clusters of\\ bronchial cells.\\ b) Columnar cell with\\ round nucleus\\ and cytoplasm.\end{tabular} & \begin{tabular}[c]{@{}l@{}}a) Polygonal /round/fibre\\ like cells.\\ b) Abundant\\ dense\\ and\\ smooth\\ cytoplasm filled with keratin.\\ c) Small hyperchromatic\\ nucleus.\\ d) Inconspicuous nucleolous.\end{tabular} & \begin{tabular}[c]{@{}l@{}}a) Accuracy of sputum cytology is\\ low and susceptible\\ to location in malignancy\\ due to less epithelial cells.\\ b) Fails to subclassify malignancies in\\ adenocarcinoma\end{tabular} \\ \hline

			14 & Salivary gland & \begin{tabular}[c]{@{}c@{}}To detect salivary gland\\ lesions.\end{tabular} & FNA & \begin{tabular}[c]{@{}l@{}}Se : \textgreater{}90\%\\ Sp : \textgreater{}90\%\end{tabular} & \begin{tabular}[c]{@{}l@{}}a) Intercalated/flat sheets\\ and tubules of duct cells\\ arranged uniform, smallshaped\\ consisting of\\ dense scant cytoplasm and\\ uniform nuclei.\\ b) Sparsely cellular with round\\ clusters of serous cells.\end{tabular} & \begin{tabular}[c]{@{}l@{}}a) Polygonal cells\\ b) Abundant granular or\\ vacuolated cytoplasm.\\ c) Presence of prominent\\ nucleoli. d) Background\\ necrosis prominent.\\ (See Fig.6)\end{tabular} & \begin{tabular}[c]{@{}l@{}}a) Malignancies cannot be\\ distinguished\end{tabular} \\ \hline

			15 & Soft tissue masses & \begin{tabular}[c]{@{}c@{}}Screening tool for\\ soft tissue masses\end{tabular} & FNA & \begin{tabular}[c]{@{}l@{}}Se :95\%\\ Sp :54\%-98\%\\ Acc : \textgreater{}90\%\\ F.P : 0\%-5\%\\ F.N : 2\%-15\%\end{tabular} & \begin{tabular}[c]{@{}l@{}}a) Small bland nuclei without atypia.\\ b)Univacuolated\\ nuclei of uniform size.\end{tabular} & \begin{tabular}[c]{@{}l@{}}a) Hypercullular\\ b) Pleomorphism.\\ c) Mitoses and background\\ necrosis present\end{tabular} & \begin{tabular}[c]{@{}l@{}}a) Intermediatory conditions between\\ benign and\\ malignant are tough\\ to determine.\end{tabular} \\ \hline

			16 & Thyroid cytology & \begin{tabular}[c]{@{}c@{}}To indicate presence of\\ thyroid nodule\end{tabular} & FNA & \begin{tabular}[c]{@{}l@{}}F.P : 1\% - 3\%\\ F.N : \textless{}1\%\end{tabular} & \begin{tabular}[c]{@{}l@{}}a) Sparsely cellular.\\ b) uniform and evenly\\ spaced follicular cells.\\ c) Coarse\\ chromatin\\ pattern.\\ (See Fig.4.i.a)\end{tabular} & \begin{tabular}[c]{@{}l@{}}a) Enlarged nucleus.\\ b) Nucleus\\ oval or elongated in shape with\\ irregular contours, pale in color.\\ Overlapped nuclei.\\ (See Fig.4.i.b)\end{tabular} & a) Sampling error \\ \hline

			17 & Urine bladder washing & To detect bladder cancer & \multicolumn{1}{l|}{\begin{tabular}[c]{@{}l@{}}Depends on urine specimen\\ type :\\ Voided, catheterized bladder\\ washing, illeal loop\end{tabular}} & \begin{tabular}[c]{@{}l@{}}Catherized\\ urine :\\ Sp :95\% -100\%\\ Se : 75\%\\ F.P :1.3\%-15\%\end{tabular} & \begin{tabular}[c]{@{}l@{}}a) Presence of crystals\\ b) Enlarged nucleus\\ with prominent\\ nucleolus.\\ c) Coarse,vacuolated\\ cytoplasm.\\ (See Fig.5.i.a)\end{tabular} & \begin{tabular}[c]{@{}l@{}}a) Round dark nuclei\\ b) Background\\ necrosis\\ (See Fig.5.i.b)\end{tabular} & \begin{tabular}[c]{@{}l@{}}a) Distinguishing characteristics\\ between high grade\\ lesions from low\\ grade lesions are ambiguous.\end{tabular} \\ \hline

		\end{tabular}%

	}

\end{table}

\clearpage

\begin{figure}[t]
	\centering
	\includegraphics[width=1\linewidth]{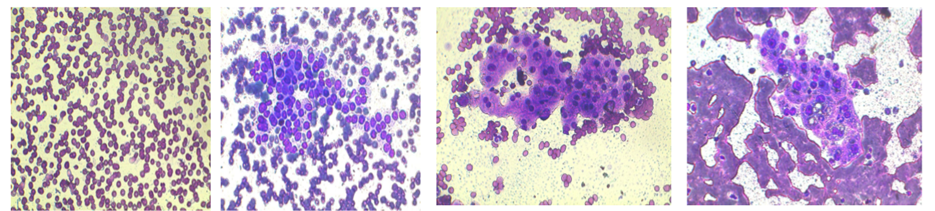}
	\caption{{Sample images of (i)Thyroid cytology(40X magnification) (a) Benign case, (b) Malignant case and (ii)Liver cytology(40X magnification) (a) Benign case, (b) Malignant case}}
	\label{fig:im1}
\end{figure}

\begin{figure}[t]
	\centering
	\includegraphics[width=1\linewidth]{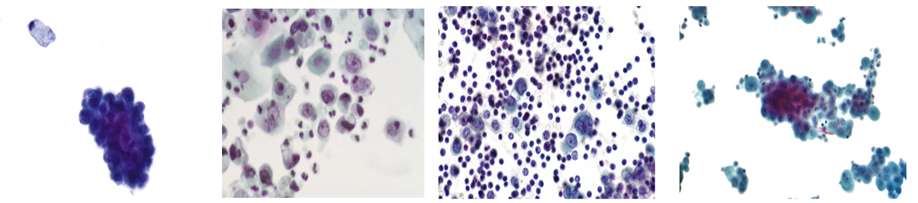}
	\caption{{ Sample images of (i)Urine cytology (a)Benign case\cite{im3}(40X magnification), (b) Urothelial carcinoma\cite{im31}(40X magnification) and (ii)Pleural fluid cytology (a) Benign mesothelial cells (pleural fluid) \cite{im4}(40X magnification), (b) Mesothelioma cytology \cite{im41}(40X magnification)}}
	\label{fig:im3}
\end{figure}

\begin{figure}[t]
	\centering
	\includegraphics[width=0.8\linewidth]{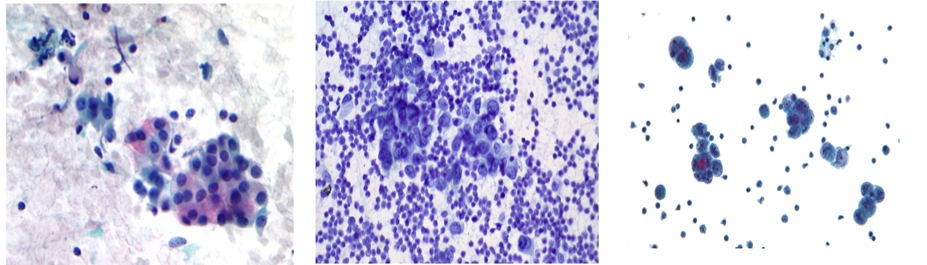}
	\caption{Sample images of (i)Salivary gland cytology(Mucoepidermoid carcinoma)\cite{im5}(40X magnification), (ii) Lymph nodes(Adenocarcinoma in Hilar Lymph Node)\cite{im6}(40X magnification),(iii) Ovarian lesions cytology(Serous carcinoma cytology)\cite{im7}(40X magnification}
	\label{fig:im5}
\end{figure}

\subsection{\textbf{Cervical cell cytology}}

According to the latest  report by WHO \cite{who}, cervical cancer ranks fourth among cancers frequently found in women. It originates
from woman's cervix and  invades other parts of the body later. In earlier stages, there is hardly any symptom. In advanced stage, symptoms like  abdominal pain, vaginal bleeding, abnormal watery discharge,
bleeding after menopause etc. are observed. The origin of cervical cancer is human papilloma virus (HPV) that causes abnormal growth of
cells of cervix( cervical dysplasia). There are mainly four types of cervical cancer such as:

a) \textbf{Squamous cell carcinomas:} It is the most commonly known form of malignant tumor of cervix and accounts for nearly 85\% of the overall cervical cancer in women of age between 40 and 55 years. These cells are lying on the outer surface of the  cervix and are thin and flat shaped.
b) \textbf{Adenocarcinomas:} It originates in glandular cells that line the upper area of cervix and
accounts to 10\% of the overall cases of cervical cancer. {Mean age of patients is 55 years.}
c) \textbf{Adenosquamous carcinomas:} It consists of both the squamous and
glandular cells and accounts for nearly 2-3\% of cervical cancers.
d) \textbf{Small cell carcinomas :} Generally very aggressive in nature and rarely found (2-3\%) which
corresponds to stage IV of the cancer. Its cytomorphology exhibits a nesting pattern and often
arranged in sheets.

\subsubsection{\textit{Modalities of cervical specimen collection}}

Cervical smears are normally obtained using a spatula and brush and {at so falls} under brush
cytology. A plastic spatula is rotated 360 degrees and the samples are exfoliated from ectocervix and endocervix. The sample is smeared on one half of the slide. The slide is then spray fixed immediately
to avoid any air drying artifact  to prevent distortion of cells. This technique is known as Papanicolaou test or Pap smear test.

\subsubsection{\textit{Differential characteristics associated with cytomorphology of malignant cells}}

{Malignant cells of cervix are generally }found in clusters of small cells with scant cytoplasm as shown in Fig.\ref{fig:7}. Nuclear
hyperchromasia and nuclear membrane irregularity are also commonly observed in malignant cells.
These characteristics are crucial in categorizing the specimen into benign and malignant using both manual
and automated analysis.
\begin{figure}[h]
	\centering
	\includegraphics[width=1\linewidth]{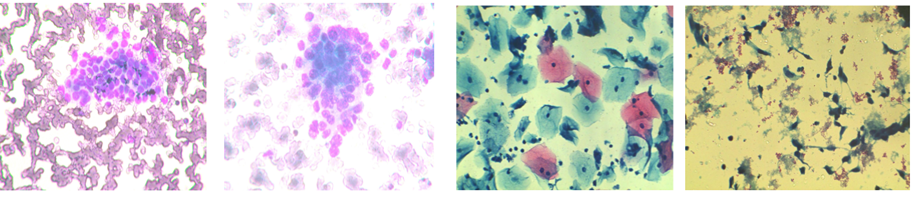}
	\caption{{Sample images of (i) Breast cytology(40X magnification) (a)Benign case(b)Malignant case and (ii)Cervical Cytology(40X magnification)(a)Benign case(b) Malignant case}}
	\label{fig:7}
\end{figure}
\subsubsection{\textit{Significant works on automated screening of cervical cell}}

Malignant nuclei differ significantly from benign nuclei in shape, size and textural pattern. Thus, nuclei have sufficient diagnostic information embedded in it and hence possess
enormous discriminative power to classify various stages of cancer {\cite{cervix}}. {Thus, accurate segmentation
	of nuclei is of prime importance for the researchers, working on cytological images.} We
categorize the review of research works based on two approaches as stated previously: i) Segmentation
based approach, ii) Segmentation free approach.
\\
\textbf{i)\textit{Segmentation Based Approach}}

A good segmentation algorithm is the necessary footstep for extracting nuclei discriminating features
and hence classify the image based on the extracted features. Various segmentation approaches to classify cervical
cytology images are discussed in this section.
\\
\textit{\textbf{Thresholding based:}} This is one of the most common and simplest techniques for extraction of foreground pixels. Walker
\cite{Walker1997AdaptiveCytology} cited  some morphological operations based on octagonal structuring element.Coarse segmentation of gray-scale images was done using  Global thresholding. A closing
morphological transform having a structural element of size smaller than the smallest nucleus removed the cytoplasm region. The nuclear
heterogeneity was adjusted by an opening operator of that size. Still, it could not be fully
automated. Zhang et al.
\cite{Zhang2011ACytology} proposed  adaptive thresholding method where the RGB image was converted into HSV image.  V-channel was extracted for histogram stretching. In the preprocessing stage  median filtering was done to remove the noise in the image. Adaptive
thresholding well mitigated the issue of non-uniform illumination.} But choosing the size of the
neighborhood  and an offset value($\theta=15$) in an optimized way, was the major
hurdle to segment clustered nuclei. Therefore, in order to segment
overlapped nuclei,  concave point based method  was applied with significant reduction in computational time.
\\
\textit{\textbf{Region based:}} 
Region based segmentation technique seeks to divide the image into sub images based on some
homogeneity conditions. Mat-Isa et al. \cite{Mat-Isa2005AutomatedAlgorithm} designed a combination of moving k-Means algorithm and
modified seed based region growing (MSBRG) algorithm to extract the region of interest
(ROI) from the image. Each pixel was assigned to the nearest cluster point using
a suitable threshold by Moving k-Means algorithm. Based on calculated thresholds and dynamic seed points MSBRG was used to delineate the
nucleus and cytoplasm edges.Thin prep
images are generally afflicted by various factors like low contrast, haziness, noise resulting in increased false
negative cases. Mustafa et al.
\cite{Mustafa2007ColourImages} proposed to enhance the contrast of thin prep images
via three different methods viz. linear contrast, nonlinear bright  and dark contrast. An
MSBRG method was used to segment nuclei. The later two contrast enhancement methods enhanced
the images by increasing the contrast of the bright areas and dark areas respectively.
In another work, Mat-Isa et al. \cite{Mat-Isa2008AnSystem} proposed a CAD based system based on region-growing-based features extraction (RGBFE) technique. It comprised of  a) An automatic feature extraction system to extract features like
nucleus size, grey level intensity of nucleus , size of the cytoplasm  and
\begin{table*}[h]
	\caption{\label{tab1}{A brief overview of different works in Cervical cancer  cytology identification where Accuracy (Acc), Jaccard Index(JI),
			Specificity (Sp), Sensitivity(Se) and Dice Similarity Coefficient (DSC) are represented in their corresponding bucket given in abbreviated form.}}
	\centering
	\label{tab:2}
	\resizebox{\textwidth}{!}{%
		\begin{tabular}{|c|c|l|c|l|l|}
			\hline
			\textbf{SL\#} & \textbf{Author’s name} & \multicolumn{1}{c|}{\textbf{\begin{tabular}[c]{@{}c@{}}Name/Source(N/S) and\\ Number of samples(Size)\\ in the Dataset\end{tabular}}} & \textbf{Method for segmentation image} & \multicolumn{1}{c|}{\textbf{\begin{tabular}[c]{@{}c@{}}Features of Nucleus\\ and cytoplasm and\\ classifier used\end{tabular}}} & \multicolumn{1}{c|}{\textbf{\begin{tabular}[c]{@{}c@{}}Quantitative results and\\ Findings (Accuracy (Acc),\\ Jacard Index(JI), Specificity\\ (Sp), Sensitivity(Se), Dice\\ Similarity Coefficient (DSC))\end{tabular}}} \\ \hline
			1 & \begin{tabular}[c]{@{}c@{}} %%Holmquist\\ 1978
				Holmquist et al.\\ (1978)\cite{Holmquist1978ComputerClassification.}\end{tabular} & \begin{tabular}[c]{@{}l@{}}N/S: University Hospital\\ (B.S.), Uppsala, Sweden\\ Size: 720 single cells of\\ eight different categories (4\\ normal and 4 malignant )\end{tabular} & \begin{tabular}[c]{@{}c@{}}Nuclei of single cells are\\ segmented using dual\\ wavelength method\end{tabular} & \begin{tabular}[c]{@{}l@{}}Features: Density, shape,\\ and texture-\\ oriented features\\ of nuclei.\\ Classifier: LDA based\\ binary classifier.\end{tabular} & \begin{tabular}[c]{@{}l@{}}Acc: 97\% in binary class\\ Findings: Algorithm does not\\ deal with superimposed cells.\end{tabular} \\ \hline
			2 & \begin{tabular}[c]{@{}c@{}} %%Mat-Isa et al. 2008\\
				Mat-Isa et al.\\(2008)
				\cite{Mat-Isa2008AnSystem}\end{tabular} & \begin{tabular}[c]{@{}l@{}}N/S: Kota Bharu Hospital\\ and Universiti Sains Malaysia\\ Hospital, Malaysia\\ Size: 550 images of three\\ class\end{tabular} & \begin{tabular}[c]{@{}c@{}}Modified seed-based \\region-growing algorithm \end{tabular} & \begin{tabular}[c]{@{}l@{}}Features: Nucleus size,\\ nucleus grey level,\\ cytoplasm size\\ and cytoplasm grey level,\\ Region-growing based\\ features extraction .\\ Classifier: H$^{2}$MLP.\end{tabular} & \begin{tabular}[c]{@{}l@{}}Acc: 97.5\\ Sp: 100\\ Se: 96.67\\ Findings: Algorithm does not\\ deal with superimposed cells.\end{tabular} \\ \hline
			3 & \begin{tabular}[c]{@{}c@{}}%%Chankong et\\ al. 2009
				Chankong et al. \\(2009)
				\cite{Chankong2009CervicalEngineering}\end{tabular} & \begin{tabular}[c]{@{}l@{}}N/S: ERUDIT Dataset.\\ Size: 276 single\\ cell images. 138 normal cells\\ ( 63 superficial cells, 75\\ intermediate cells), and 138\\ abnormal cells ( 34 light\\ dysplastic cells, 34 moder-\\ ate dysplastic cells, 35\\ severe dysplastic cells and 35\\ carcinoma in situ cells).\end{tabular} & Fourier Transform & \begin{tabular}[c]{@{}l@{}}Features: Mean,\\ variance, and entropy.\\ Classifier: Bayesian\\ classifier,\\ linear\\ discriminant analysis\\ (LDA), KNN, ANN,\\ and SVM\end{tabular} & \begin{tabular}[c]{@{}l@{}}Acc: 92.65 (Highest SVM)\\ False positives:7.35\%\\ False negatives: 7.35\%\\ Findings: Experimented with\\ single cells. Overlapping cells\\ which is the practical scenario\\ are not considered.\end{tabular} \\ \hline
			4 & \begin{tabular}[c]{@{}c@{}}Li et al.\\ 2012
				\cite{Li2012CytoplasmSnake}\end{tabular} & \begin{tabular}[c]{@{}l@{}}N/S: Herlev dataset,\\ Size: 917 images\end{tabular} & \begin{tabular}[c]{@{}c@{}}Spatial K-means clustering\\ followed by RGVF Snake\\ incorporating a stack-based\\ refinement using edge map\\ .\end{tabular} & \begin{tabular}[c]{@{}l@{}}Classifier:\\ interpretation. visual\end{tabular} & \begin{tabular}[c]{@{}l@{}}Zijdenbos similarity index: 94\%\\ Findings: only segmentation of\\ isolated cells. Overlapping of\\ cells not considered.\end{tabular} \\ \hline
			5 &  Genctav et al.\cite{Genctav2012UnsupervisedImages} & \begin{tabular}[c]{@{}l@{}} Herlev data set, \\Size:917 images of Pap cell  \\Hacettepe data set, \\Size:82 Pap test images \end{tabular} & Non-parametric hierarchical segmentation. & \begin{tabular}[c]{@{}l@{}}Features: 12 features\\ of cells including\\ spectral and shape.\\ Classifier:\\ spectral, shape and gradient\\ information of nucleus\\ for classification.\end{tabular} & \begin{tabular}[c]{@{}l@{}}Correct classification rate is 96 \% \\ Findings: Parameter free segmentation. \\ Clusters of overlapped regions\\ not addressed. \end{tabular} \\ \hline
			6 & \begin{tabular}[c]{@{}c@{}}Zhang et al.\\ (2014)
				\cite{ZHANG2014369}\end{tabular} & \begin{tabular}[c]{@{}l@{}}N/S: Shenzhen Sixth\\ People's Hospital\\ Huazhong University of\\ Science and Technology\\ Union Shenzhen Hospital,\\ Shenzhen, China, 2010.\\ Size: 51 images (21 slides ).\end{tabular} & \begin{tabular}[c]{@{}c@{}}Local graph cut and global\\ graph cut to delineate\\ cytoplasm and nucleus.\\ Concave point\\ based algorithms\\ to segment\\ overlapping nucleus.\end{tabular} & \begin{tabular}[c]{@{}l@{}}Features: nucleus and\\ cytoplasm texture,\\ shape and size.\\ Classifier:\\ Visual interpretation(only\\ segmentation is\\ reported)\end{tabular} & \begin{tabular}[c]{@{}l@{}}Acc: 93\% for segmentation\\ purpose\\ Findings: Touching nuclei are\\ segmented.\\ Poor contrast images \\are underestimated.\end{tabular} \\ \hline
			7 & \begin{tabular}[c]{@{}c@{}}Nosrati and Hamarneh \\(2015)
				\cite{7163846}\end{tabular} & \begin{tabular}[c]{@{}l@{}}N/S: ISBI Overlapping\\ Cervical Cytology Image\\ Segmentation Challenge.\\ Size: 135 images (45\\ images for training and 90\\ images for testing).\end{tabular} & \begin{tabular}[c]{@{}c@{}}Star shape prior and Voronoi\\ energy term.\end{tabular} & \begin{tabular}[c]{@{}l@{}}Features:\\ (HOG) features are\\ used for training the\\ nucleus to cytoplasm\\ ratio.\\ Classifier:\\ Visual interpretation.\end{tabular} & \begin{tabular}[c]{@{}l@{}}DSC: 0.88\\ Findings:\\ In multiple overlapped regions\\ the boundary of each cell is little\\ ambiguous. Time to segment\\ each image is 6.6 sec. (approx.).\end{tabular} \\ \hline
			8 & \begin{tabular}[c]{@{}c@{}} 
				Mariarputham \\and Stephen\\ (2015)\cite{Mariarputham2015NominatedClassification}\end{tabular} & \begin{tabular}[c]{@{}l@{}}N/S: Herlev dataset\\ Size: 917images\end{tabular} & \begin{tabular}[c]{@{}c@{}}Image inversion, binarization,\\ morphological closing with \\ structuring element of size 5 and\\ filling operation.\end{tabular} & \begin{tabular}[c]{@{}l@{}}Features: 24 features\\ including nucleus\\ related, moment,\\ texture etc.\\ Classifiers: SVM and\\ Neural Network .\end{tabular} & \begin{tabular}[c]{@{}l@{}}Acc: 81.85\% (best SVM) for\\ seven class problem on 10 fold\\ cross validation.\\Findings: Dill with only\\ single cell images.\end{tabular} \\ \hline
			9 & \begin{tabular}[c]{@{}c@{}}%%Zhang\\ et al. 2017
				Zhang et al.\\ (2017a)\cite{Zhang2017Graph-basedCytology}\end{tabular} & \begin{tabular}[c]{@{}l@{}}N/S: 1)Herlev dataset\\ 2)HandE stained manual\\ LBC Dataset\\ Size: Herlev dataset(917),\\ HEMLBC dataset(21)\end{tabular} & Graph Search based segmentation of nuclei & \begin{tabular}[c]{@{}l@{}}Assessment based on \\5--fold\\ segmentation\\ performance.\end{tabular} & \begin{tabular}[c]{@{}l@{}}On Herlev dataset(for carcinoma\\ having 150 cells) :\\ Precision:93\%$\pm$0.08,\\ Recall:91\%$\pm$0.13\\ On HEMLBC dataset(for\\ carcinoma having 150 cells) ::\\ Precision:91\%$\pm$ 0.04,\\ Recall:96$\pm$ 0.04\% \\Findings: Global optimal solution\\ achieved. Overlapped cells \\are not addressed, also accuracy\\ is highly dependent on\\ initial segmentation .\end{tabular} \\ \hline 
			10& \begin{tabular}[c]{@{}c@{}}%%Zhang\\ et al. 2017
				Zhang et al.\\ (2017b)\cite{Zhang2017DeepPap:Classification}\end{tabular} & \begin{tabular}[c]{@{}l@{}}N/S: 1)Herlev dataset\\ 2)HandE stained manual\\ LBC Dataset\\ Size: Herlev dataset(917),\\ HEMLBC dataset(21)\end{tabular}  & No segmentation & \begin{tabular}[c]{@{}l@{}}Used previously\\ trained CNN (trained\\ on ImageNet)\\ 5 convolution layer and 3\\ forward connection\\ layer.\end{tabular} & \begin{tabular}[c]{@{}l@{}}On Herlev:\\ Acc: 98.3\%,\ Sp: 98.3\%\\ Se: 98.2\%\\ On HEMLBC:\\ Acc: 98.6\%,\\ Sp: 99.0 \%\\ Se: 98.3\% \\Findings: Multiple and overlapper\\ cells not considered.\\ Misclassification rate\\ is higher.\end{tabular} \\ \hline
		\end{tabular}%
	}
\end{table*} \clearpage
 grey level intensity of cytoplasm  using region growing method
and b) A diagnostic system  was designed using hierarchical hybrid multilayered perceptron
(HHMLP). An accuracy of 97.50\% was reported on their dataset  consisting of 550 reported cases "(211 normal, 143 {Low-grade Squamous Intra-epithelial Lesion} (LSIL) and 196 {High-grade Squamous Intra-epithelial Neoplasia} (HSIL) cases)".
\\
\textit{\textbf{Contour based:}}
Active contour or snake model (ACM) seek
to segment the image using the initial contour information. But ACM fails for poor quality images and is also sensitive to the detection
of initial contour. Snake model is one of the popularly used contour models for segmention
nuclei from cytoplasm and background. Yang-Mao et al. \cite{Yang-Mao2008EdgeImages} proposed an edge enhancement nucleus and
cytoplast contour detector (EENCC) algorithm involving five steps: trim-meaning filter, bi-group
enhancer,  computation of gradient, MVD (Maximum Valued Difference), and contour extraction. EENCC  performed better than snake models with least relative distance error (RDE) of 0.288. Harand et al. \cite{Harandi2010AnThinPrep} adopted a geometric active contour algorithm 
followed by Sauvola thresholding and FIFO queue structure % Please add the following required packages to your document preamble:
to identify boundary of nucleus within each cell. The contour obtained in low resolution is taken as the input. It showed better results compared to the EENCC 
detector.Guan et al. \cite{Guan2015AccurateSO} attempted to segment partially overlapping
cervical cells in high resolution images. A dynamic sparse contour algorithm was used to detect weak contour points in the nucleus. Then GVF (Gradient Vector Flow) snake model was applied to find out the exact contour of cell. However, the method could not handle more than two overlapped cells.
Moreover, its performance degrades with low resolution images.

\textit{\textbf{Deformable Contour model based:}} 
The Active Contour Model (ACM) pioneered by Xu et al.\cite{Xu1998SnakesFlow}
is extensively used to extract the objects with irregular boundaries, contour with noisy interior and contour with small breaches.{  Tsai et al.\cite{Tsai:2008:NCC:1371261.1371379} 
developed a three--staged cytoplasm
and Nucleus Contour (CNC) detection technique for segmentation of nuclei and cytoplasm. A bi-group enhancer, in first stage was used to sharpen the contour of nucleus by increasing the
contrast between nucleus and cytoplasm regions. 
Cytoplasm contour detection using K-means clustering was applied in second stage. In the last stage, contour of the nucleus was
detected by applying Maximal Color Difference (MCD) method.} But a prior denoising by means of
median filter is a prerequisite to acquire a satisfactory result. Li et al. \cite{Li2012CytoplasmSnake} used priory spatial K-means clustering algorithm to segment the  image into background, nucleus, and cytoplasm for free lying cells.
A radiating gradient vector flow (RGVF) based snake algorithm was applied to get fine segmentation. RGVF involved a stack-based refinement and an edge map based technique to
identify the unclear and incomplete edges, but eventually  failed to segment  overlapping cells.
Furthermore, the performance was highly dependent on accuracy of initial contour extraction. Bergmeir et al.\cite{Bergmeir2012SegmentationFramework}
proposed a web-based software module with a prior noise removal technique using mean-shift and a median
filtering. A randomized Hough transform was used to detect
candidate nucleus that was segmented by level set methods. The experiment was done on 207 cervical images with an F-measure of 96.15\% and true positive rate of 95.63\%. But in the proposed system
ROI was selected manually by cytotechnologists.Lu et al. \cite{10.1007/978-3-642-40811-3_57} proposed the segmentation of non-overlapped nuclei using MSER (Maximally
Stable Extremal Regions) technique where cell clumps were taken as input. To segment overlapped
cells they used joint level set optimization. But, considering the shape of the nucleus as ellipse to model
the segmentation problem, the accuracy level was compromosed to some extent. {But
the cytoplasm boundary for low contrast images could not be properly delineated also the system should be validated with true
dataset.} To overcome this problem, Nosrati et al.\cite{7163846} assumed a star shape prior instead of ellipse, for
better approximation of the nuclei shape,  which was encoded  in a vibrational framework having
directional derivatives. To address overlapped cells, they included Voronoi
energy term, which bounded the amount of overlapping in adjacent cytoplasm. 
\\
\textit{\textbf{Texture based:}}  
Texture is an intrinsic property of an image which can be used to extract useful contents
from an image. So, researches are motivated towards texture analysis for classification of pap-smear
images.{ Among seven different sets of texture feature, LBP{(Local Binary Pattern)} and GLCM(Gray Level Co-Occurrence
Matrix) were used for classifying pap smear cells of Herlev dataset, a standard freely available Pap
Smear dataset, consisting of 917 samples, into seven output classes using SVM(Support Vector Machine) and NN(Neural Network) classifier \cite{Mariarputham2015NominatedClassification}.}
\\
\textit{\textbf{Graph based:}} 
A graph-cut approach was advised by Zhang et al. \cite{Zhang2014AutomatedCuts} to segment cervical cells of
manual liquid based cytology.  A multi-way Graph-cut approach was used to segment
cytoplasm. This technique  grossly classified the image into background (with lowest mean
intensity) and cytoplasm, nuclei, inflammatory cells and debris. For precise nuclei segmentation, they
used graph-cut approach adaptively and locally. To split touching nuclei, two concave point based
algorithms were combined. But the system could not detect cytoplasm boundary. Also, nucleus having
poor contrast {could not be} delineated completely. A graph search based cervical cell segmentation
methodology was presented by Zhang et al. \cite{Zhang2017Graph-basedCytology}. Initially they defined the center point and the size of the
nucleus. For segmentation of nucleus boundary they used graph search based method where the shape
information was plotted as a graph (Cartesian to polar plot) by designing a cost function based on edge
term and region term. Optimal path of the graph was dictated by dynamic programming. To obtain
closed contour of nucleus in Cartesian coordinate system, reverse mapping was done. This work was
tested on two standard datasets: Herlev and HEMLBC dataset. This method showed fast optimization
process but the accuracy depends greatly on closeness of initially detected nucleus center point to the
actual one.
\\
\textit{\textbf{Clustering based:}} { Kim et al.\cite{Kim2007NucleusAlgorithm} suggested a Fuzzy C-Means (FCM) based
segmentation method on   uterine cervical images in  HSI colour space.} {A patch
based FCM clustering technique was proposed by Chankong et al.\cite{article} for
segmentation of nucleus, cytoplasm and background. Six nuclei features were extracted to classify cervical cells using  five classifiers: Bayesian
classifier, Linear discriminant analysis (LDA), KNN, SVM, ANN. They experimented on ERUDIT Pap
smear dataset, LCH Dataset and Herlev dataset. For a two-class problem, ANN gave best recognition
accuracy of 97.83\%, 97.00\% and 99.27\% on three datasets respectively.} Zhao et al.\cite{Zhao2016AutomaticMRF} applied  Markov Random Field (MRF)based segmentation model after  getting super pixels  from cervical images.   The gap search algorithm with reduced time complexity used further based on the labelling of nucleus, cytoplasm and other components of the cell.
\\
{\textbf{Deep Learning based:}}
Deep learning based framework is one of the latest trends in many applications and being extensively used in segmentation of cytology images. {Song et al.\cite{song2014deep} developed a CNN (Convolutional Neural Network) based segmentation framework combined with a superpixel based technique for segmentation of nucleus. The gaussian noise generated during image acquisition process was removed by trimmed meaning filter. The cytoplasmic mask was extracted by high-dimensional Otsu thresholding method. A superpixel based SLIC (Simple Linear Iterative Clustering) segmentation was applied on masked images. To extract  features of superpixels, CNN was applied.  Coarse nuclei segmentation was done to reduce overlapping of inflammatory cells. A new template was constructed to improve segmentation. The nuclei region detection accuracy was obtained as 94.50 \%  using CNN.} 
Liu et al.\cite{liu2018automatic} proposed an automated image segmentation method by Mask-RCNN (Mask- Regional Convolutional Neural Network) and LFCCRF (Local Fully Connected Conditional Random Field).The Mask-RCNN, which was constructed by residual neural network based FPN (Feature Pyramid Network),  was modified depending on  cervical nuclei images. The pyramid feature maps were extracted by Mask-RCNN through pixel level information of the nuclei. The Mask-RCNN based coarse segmentation of nuclei were performed and ROI  was obtained by increasing the bounding box. For further refinement LFCCRF was used with  intensity and position information of all pixels in that region. Phoulady et al.\cite{phoulady2018new} proposed a CNN based nucleus detection technique using iterative thresholding  method to  find the seed point of nuclei. CNN was trained with the patches of  nuclei. This approach showed a precision, recall and F-score of  0.861,0.895 and 0.878 respectively on the cervical cytology dataset 'CERVIX93'.
\\

\textit{\textbf{Other approaches:}} \cite{7482168} studied the influence of ICM (Intersecting Cortical Model)where the parameters were optimized using Particle Swarm Optimization (PSO). Several other algorithms like Otsu thresholding, Expectation Maximization, Region Growing and
Fuzzy C-Means clustering were studied in the experiment with 250 cervical images. ICM was reported with a highest PSNR of 62.946 dB. An unsupervised segmentation of Pap smear cells was proposed by  Happy et al.\cite{DBLP:journals/corr/HappyCS15} in two stages. The cells were divided based on
homogeneity and circularity using multi-scale hierarchical segmentation. {The nucleus and cytoplasm was classified using a binary classifier.} They segmented overlapping  nuclei using extended depth of field (EDF) images from different focal planes with
various degrees of overlap. Modified Otsu using class prior probability was used to select suitable threshold 
to segment nucleus. The work was validated on ISBI
2015 challenge dataset with a DICE score of 0.86 and a TP rate 0.88. Lakshmi et al.\cite{AnnaLakshmi2017AutomatedICM} applied Haar wavelet to nullify the effect of uneven staining
in the pap-smear images followed by adaptive median filter to remove noise. ICM and cuckoo search algorithm jointly used to segment the nucleus in an optimized way. however the initial parameters
were chosen manually.  Iliyasu et al.
\cite{Iliyasu2017ADetection} investigated a hybrid approach of feature set selection. They combined quantum particle swarm optimization (QPSO) algorithm with Fuzzy k-
nearest neighbours (Fuzzy KNN) algorithm to scale down  17 features relating color, geometry
and texture of nucleus to 7 features. Selected features prior to hybrid approach based classification  showed greater classification accuracy compared to All--feature--Based approach
(no feature selection).
{A cellular neural network based cervical cancer detection method using customized template was advocated by Abdullah et al. {\cite{abdullah2019cervical}}.  Three templates were used such as 1) modified blue channel extraction  template for identification of nuclei from the cell 2)   modified contrast enhancement  template for identifying nuclei with finer details from  modified image 3)  modified hollow concave  template for  removing noisy background. A recognition accuracy of  90.77\% and 85.54\% for cancerous and non-cancerous cells respectively.}\\
\textbf{\textit{ii)Segmentation Free Approach:}}
Recently with the onset of deep learning based techniques, deep convolutional networks {can be applied directly to classify pap-smear cells} without requiring prior segmentation. Most of the researchers used transfer learning or fusion of different
CNNs in their works. In the paper \cite{Bora2016PapNetwork},  VGG net-16, a common CNN architecture
for feature extraction,  was used for classification using Least Square Support Vector. 
Zhang et al.\cite{Zhang2017DeepPap:Classification} introduced DeepPap, a CNN architecture using the concept of
transfer learning. A pre-train architecture of image net for pap smear was used for image classification. They
observed a recognition accuracy of 98.3\% and 98.6\% on Herlev and HEMLBC dataset
respectively using five fold cross validation.  Hyeon et al.
\cite{7881741} used VGG-16 net, a pre--
trained CNN architecture for feature extraction. The extracted features were used to classify using SVM. 71,344 Pap smear images were collected from Seegene Medical
Foundation to form a dataset consisting of 8,373 samples of normal class and 8,373
samples of cancerous or abnormal class. A recognition accuracy of 78\% and 20\% was observed on the test samples of the dataset.
Some major segmentation techniques used in cervical cytology are shown in Table \ref{tab:3}.
\\
\textbf{\textit{Classification:}}
\\
After segmentation of nucleus of Pap smear images, different nuclei centric
features are extracted to classify pap smear images into two classes
Benign and Malignant. Malignant or cancerous are further classified into different
categories based on their stage.A single classifier or combination or ensemble of
different classifiers are used. Among different classifiers binary classification method \cite{Holmquist1978ComputerClassification.}, Artificial Neural Network (ANN) and its modifications   \cite{Mehdi1994AScreening,Mat-Isa2008AnSystem,article} Support vector machine (SVM) \cite{Zhang2004CervicalScreening, Orozco-Monteagudo2012CombinedExtraction},
Fuzzy C Means (FCM) \cite{Kim2007NucleusAlgorithm,Plissiti:2011:ADC:2222949.2223623}
etc. are popularly used.  Deep learning techniques
\cite{Zhang2017DeepPap:Classification,10.1007/978-3-319-60964-5_23} etc.are recently introduced for classification of cervical cells.
\\
\textbf{\textit{Binary Classification:}} %%
{Holmquist et al.\cite{Holmquist1978ComputerClassification.} developed a binary classification framework on Linear Discriminant Analysis (LDA) to
	distinguish cervical images using density, shape, and texture-oriented features 
	of  nuclei.}
\\
\textbf{\textit{Hierarchical approach:}} %% 
Mehdi et al.\cite{Mehdi1994AScreening} introduced hierarchical approach to classify images into mild, moderate and severe
cells using ANN with back propagation algorithm. Ramli et al.
\cite{Ramli2004DiagnosisNetwork} approached with a non-linear hybrid
multi-layered perception(HMLP) model using least square algorithm to classify cervical cells into normal and
low-grade squamous epithelium. %% 
Mat-Isa  et al.\cite{Mat-Isa2008AnSystem} also recommended an HMLP network to classify single cell images using different feature
informations such as nucleus size, nucleus grey level, cytoplasm size and cytoplasm grey
level etc. 
\\
\textbf{\textit{Support Vector Machine({SVM}):}} %% 
Zhang et al.\cite{Zhang2004CervicalScreening} used SVM to detect cancerous cells from multispectral
pap smear images.  
During feature selection Sequential Backward Selection (SBS) was applied.
They
obtained better performance after using SVM based screening with smaller number of features.Monteagudo et al.\cite{Orozco-Monteagudo2012CombinedExtraction} proposed a combination of SVM and waterfall algorithm also classify cancerous and normal specimens.
\\
\textbf{\textit{Fuzzy C-Means(FCM):}} %%
Kim et al.\cite{Kim2007NucleusAlgorithm} suggested FCM based nucleus recognition techniques to distinguish
normal and abnormal cells in HSI images. Different nucleus specific information such as
circumference of nucleus, different ratios from nucleus and cytoplasm, degree of roundness, etc. %%
\cite{Plissiti:2011:ADC:2222949.2223623} were used to detect true nuclei points using  different extrema points in the
images. FCM  was preferred over SVM for classifying the normal and
abnormal cells due to  limited data set.
\\
\textbf{\textit{Convolutional Neural Network:}} Song et al.\cite{Song2015AccuratePartitioning} 
proposed  multi-scale convolutional network (MSCN) and graph based partitioning technique to segment touching nuclei.  MSCN  was used for initial
segmentation of nuclei extracting multiscale feature vectors. It was
fine segmented using graph partitioning to get accurate result. When compared to raw pixel
segmentation, superpixel based segmentation of cytoplasm and nucleus gave an improved accuracy of 5.06\%
and 2.06\% respectively. Braz et al.\cite{Braz2017NucleiLearning} proposed another
CNN based technique for detection of nucleus in pap smear images. Gautam et al.
\cite{unknownc} proposed a  cell nuclei detection technique using patch based CNN.
The segmented nuclei were classified using transfer learning on AlexNet. They also proposed a
decision tree based technique. A recognition accuracy of 99.3\% in two-
class problem and 93.75\% on a seven class problem was reported on the same dataset. Wu  et al.{\cite{wu2018automatic} proposed AlexNet based Deep convolution neural network model which  was trained using three fold cross validation.For data augmentation flipping and rotation techniques were done. Training and testing process was done on both original RGB and augmented image datasets. Each dataset consisted of  keratinizing squamous, non-keratinizing squamous and
	basaloid squamous.  In different groups of datasets they reported recognition accuracy of  93.33\% and 89.48\%  for original  and augmented images respectively.} Meiquan et al. {\cite{meiquan2018cervical} proposed an automated cell detection technique using  a faster R-CNN  method. Using Resnet-101 the features were extracted. In testing phase, the data were enhanced by rotating the  patches  at $90^\circ$, $180^\circ$, $270^\circ$. A recognition accuracy of 0.91, 0.78, 0.70 was reported on validation set, two class and four class problem  respectively.}

\subsection{\textbf{BREAST cancer cytology}}

Breast cancer, though, mostly prevalent among women all over the world, occurrences found in men are
rare (cases found are mostly invasive in nature). In women, the incidence and deaths due to breast cancer
is the highest posing a global burden irrespective of all levels of modernization. Breast cancers originate
either in the lobules, or in the ducts that connect
the lobules to the nipple. Breast cancer does not show symptoms when the tumor is small enough to be felt and can be easily cured. It starts with a painless lump and slowly progresses with other symptoms  like  pain, breast heaviness, swelling, redness of the skin. Nipple abnormalities such as
spontaneous watery or bloody discharge or retraction are also common. Samples of cytology images of breast are shown in Fig.{\ref{fig:7}}. Breast cancer is generally categorized into  i) Carcinoma In Situ, ii) Invasive.

{\textbf{i)In Situ:}} Here instances of  abnormal cells that are found locally and have not spread to nearby cells. These
are  mainly of  two types: ductal carcinoma in-situ (DCIS) and lobular carcinoma in
situ (LCIS)or lobular neoplasia.
\begin{table*}[!t]
	\caption{\label{tab1}A brief overview of different works in Breast cancer cytology identification where Accuracy (Acc), Jaccard Index(JI),
		Specificity (Sp), Sensitivity(Se) and Dice Similarity Coefficient (DSC) are represented in their corresponding bucket given in
		abbreviated form.}
	\label{tab:3}
	\centering
	\resizebox{\textwidth}{!}{%
		\begin{tabular}{|l|l|l|l|l|l|}
			\hline
			\multicolumn{1}{|c|}{\textbf{SL \#}} & \multicolumn{1}{c|}{\textbf{\begin{tabular}[c]{@{}c@{}}Author’s\\ name\end{tabular}}} & \multicolumn{1}{c|}{\textbf{\begin{tabular}[c]{@{}c@{}}Name/Source(N/S) and\\ Number of samples\\ (Size) in the Dataset\end{tabular}}} & \multicolumn{1}{c|}{\textbf{\begin{tabular}[c]{@{}c@{}}Method for image\\ segmentation\end{tabular}}} & \multicolumn{1}{c|}{\textbf{Features and classifier used}} & \multicolumn{1}{c|}{\textbf{\begin{tabular}[c]{@{}c@{}}Quantitative results\\ and Findings\end{tabular}}} \\ \hline
			1 & \begin{tabular}[c]{@{}l@{}} Wolberg et al.\\ (1993)\cite{Wolberg1993BreastAnalysis.}\end{tabular} & \begin{tabular}[c]{@{}l@{}}N/S: University of Wisconsin Clinical\\ Sciences Center Highland Avenue Madison,\\ Wisconsin, USA.\\ Size: 119 images. 68 benign 51 malignant.\end{tabular} & \begin{tabular}[c]{@{}l@{}}Manual segmentation\\ followed by active\\ contour detection.\end{tabular} & \begin{tabular}[c]{@{}l@{}}Features: size, perimeter,\\ area, compactness, radial\\ variance, concavity, texture,\\ size uniformity, worst size,\\ worst shape.\\ Classifier: MSM-Tree.\end{tabular} & \begin{tabular}[c]{@{}l@{}}Acc: 90\%\\ Findings:\\ Manual segmentation\\ of nucleus boundary\\ is cumbersome and time\\ consuming\end{tabular} \\ \hline
			2 & \begin{tabular}[c]{@{}l@{}} Weyn et al. \\(1998)\cite{Weyn1998AutomatedDescription.}\end{tabular} & \begin{tabular}[c]{@{}l@{}}N/S: Antwerp University Hospital,\\ Wilrijk, Belgium.\\ Size: 83 images. 20 benign, 63 malignant\end{tabular} & \begin{tabular}[c]{@{}l@{}}Segmentation not done\end{tabular} & \begin{tabular}[c]{@{}l@{}}Features: Wavelet, Densitometric,\\ co-occurrence,\\ morphometric features.\\ Classifier: KNN\end{tabular} & \begin{tabular}[c]{@{}l@{}}Acc: 76.1\%\\ Findings: Operator\\ supervision\\ is required and also false\\ negatives present.\end{tabular} \\ \hline
			3 & \begin{tabular}[c]{@{}l@{}} Street\\ (2000a)\cite{Street2000Xcyt:Cancer}\end{tabular} & \begin{tabular}[c]{@{}l@{}}N/S: University of Wisconsin Hospitals \\ and Clinics beginning in 1984.\\ Size: 569 images. 357 benign, 212 malignant.\end{tabular} & \begin{tabular}[c]{@{}l@{}}Generalized Hough\\ transform\end{tabular} & \begin{tabular}[c]{@{}l@{}}Features: 30 numbers of\\ nuclear morphometric features.\\ Classifier: MSM-Tree\end{tabular} & \begin{tabular}[c]{@{}l@{}}Acc: 97.5\%.\\ Se: 96.7\%,\\ Sp: 98.0\%.\\ Findings: Accurate\\ closed boundaries,\\ false findings persist,\\ overlapped cells not\\ taken care of. Not\\ fully remote, requires\\ end user intervention\end{tabular} \\ \hline
			4 & \begin{tabular}[c]{@{}l@{}} Isa et al.\\(2007)\cite{Isa2007FineNetwork}\end{tabular} & \begin{tabular}[c]{@{}l@{}}N/S: Penang General Hospital and\\Hospital Universiti Malaysia,\\ Kelantan, Malaysia.\\ Size: 1300 images in 4 categories.\end{tabular} & \begin{tabular}[c]{@{}l@{}}Segmentation not done\end{tabular} & \begin{tabular}[c]{@{}l@{}}Features: 13 features of\\ different cellular and nuclei\\ properties extracted.\\ Classifier: Hybrid MLP (800\\ training and 500 test cases)\end{tabular} & \begin{tabular}[c]{@{}l@{}}Acc:100\%\\ Findings: Intelligent\\ system. where staging\\ of cancer can be\\ determined.\end{tabular} \\ \hline
			5 & \begin{tabular}[c]{@{}l@{}} Jelen et al. \\(2008b)\cite{Jelen:2008:CBC:1721542.1721549}\end{tabular} & \begin{tabular}[c]{@{}l@{}}N/S: Medical University of Wroclaw, Poland.\\ Size: 66 (intermediate malignancy),\\ 44(high malignancy)\end{tabular} & Level set method & \begin{tabular}[c]{@{}l@{}}Features: area, perimeter,\\ convexity, eccentricity and\\ one texture feature.\\ Classifier:MLP, Self-organizing\\  maps (SOM), PCA, SVM.\end{tabular} & \begin{tabular}[c]{@{}l@{}}Acc: Maximum\\ 94.24\%with SVM.\\ Findings: Degree of\\ malignancy can be\\ determined\end{tabular} \\ \hline
			6 & \begin{tabular}[c]{@{}l@{}} Malek et al. \\(2009)\cite{Malek:2009:ABC:1527246.1527265}\end{tabular} & \begin{tabular}[c]{@{}l@{}}N/S: Farhat Hached Hospital, Sousse,Tunisia.\\ Size: 200 images in test data set.\\120 benign,80 malignant.\end{tabular} & GVF Snake & \begin{tabular}[c]{@{}l@{}}Features: wavelet\\ based texture features.\\ Classifier: Fuzzy C-means\end{tabular} & \begin{tabular}[c]{@{}l@{}}Acc: 95\%\\ Findings: Developed\\ FPGA based\\ hardware system of\\ software counterpart\end{tabular} \\ \hline
			7 & \begin{tabular}[c]{@{}l@{}} Kowal et al. \\(2011)\cite{kowal2011computer}\end{tabular} & \begin{tabular}[c]{@{}l@{}}N/S: Regional Hospital in Zielona Gora, Poland.\\ Size: 500 images from 50 patients\\ (10 images/patient). 25 benign, 25 malignant.\end{tabular} & \begin{tabular}[c]{@{}l@{}}Adaptive thresholding\\ and Gaussian mixture model\\ based segmentation.\end{tabular} & \begin{tabular}[c]{@{}l@{}}Features: Different Nuclei and\\ cytoplasm centric features.\\ Classifier(s): KNN, Naive Bayes \\ classifier, DecisionTrees, Ensemble Classifier.\end{tabular} & \begin{tabular}[c]{@{}l@{}}(Maximum using KNN)\\ Acc: 98\% \\ Findings: Incorrect nuclei\\ clusters are generated\\ for lower no. of pixels.\\ No standard dataset.\end{tabular} \\ \hline
			8 & \begin{tabular}[c]{@{}l@{}} Filipczuk et al.\\ (2013a)\cite{Filipczuk2013Computer-aidedBiopsies}\end{tabular} & \begin{tabular}[c]{@{}l@{}}N/S: Regional Hospital in Zielona Gora, Poland.\\ Size: 737 images from 67 patients (cases)\end{tabular} & Circular Hough transform & \begin{tabular}[c]{@{}l@{}}Features: Nuclei and cytoplasm\\ centric features with different\\ intensities of RGB images,\\ Correlation, Energy, Homogeneity etc.\\ Classifier(s): KNN, Naive Bayes,\\ Decision Tree, SVM\end{tabular} & \begin{tabular}[c]{@{}l@{}}(Maximum using\\ SVM)\\ Acc: 98.51\% \\ Findings: Flase circles are\\ eliminated.\\No standard dataset. \end{tabular} \\ \hline
			9 & \begin{tabular}[c]{@{}l@{}} Issac et al.\\ (2013)\cite{ISSACNIWAS20132828}\end{tabular} & \begin{tabular}[c]{@{}l@{}}N/S: Pathology lab of Regional Cancer Center,\\ Thiruvananthapuram.\\ Size: 334 (benign proloferative 311(infiltrating)\end{tabular} & \begin{tabular}[c]{@{}l@{}}K-means clustering\\ technique in LAB\\ color space.\end{tabular} & \begin{tabular}[c]{@{}l@{}}Features: Textures and Nuclei\\ centric features extraction after\\ using complex Daubechies Wavelet Transform.\\ Classifier: KNN\end{tabular} & \begin{tabular}[c]{@{}l@{}}Acc: 93.9\%,\\ Se: 92.2\%\\ Sp: 95.9\%\\ Findings: Complex\\ wavelet performed\\ better than\\ real wavelet. More morp-\\ hometric features can\\ improve results.\end{tabular} \\ \hline
			10 & \begin{tabular}[c]{@{}l@{}} George et al.\\ (2014)\cite{George2014RemoteImages}\end{tabular} & \begin{tabular}[c]{@{}l@{}}N/S: Ain shams University Hospitals,Egypt\\ Size: 92 images. 45 benign, 47 malignant\\ having 11502 cell nuclei.\end{tabular} & \begin{tabular}[c]{@{}l@{}}Marker controlled\\ watershed\\ segmentation\end{tabular} & \begin{tabular}[c]{@{}l@{}}Features: 12 statistical\\ features, 10 texture features\\ and 2 intensity based features.\\ Classifier(s): SVM, LVQ, PNN, MLP\end{tabular} & \begin{tabular}[c]{@{}l@{}}Using PNN (max)\\ Se: 96.32\%\\ Sp: 94.57\%\\ Findings: Developed\\ fully automated re-\\ mote system. Gra-\\ ding of malignancy\\ not done.\end{tabular} \\ \hline
			11 & \begin{tabular}[c]{@{}l@{}} Garud et al.\\(2017)\cite{Garud2017High-MagnificationNetworks}\end{tabular} & \begin{tabular}[c]{@{}l@{}}N/S: Sub-divisional Hospital, Kharagpur, and\\ Midnapur Medical College and Hospital,\\ Midnapur, India.\\ Size: 37 samples.24 benign, 13 malignant.\end{tabular} & Manually selected ROI & \begin{tabular}[c]{@{}l@{}}Features/classifier: Images are subdivided\\ randomly to fit in GoogLeNet\\ architecture and performances\\ are reported both ROI level\\ and sample level.\end{tabular} & \begin{tabular}[c]{@{}l@{}}Acc: ROI: 80.76\%.\\ Samples: 89.71\%.\\ Findings: Nuclei ce-\\ ntric feature based\\ approach perform\\ better than\\ GoogLeNet .\end{tabular} \\ \hline
			12 & \begin{tabular}[c]{@{}l@{}} Zejmo et al.\\(2017)\cite{unknown}\end{tabular} & \begin{tabular}[c]{@{}l@{}}N/S: Regional Hospital in Zielona Gora, Poland.\\ Size: 50 patients 25 benign, 25 malignant cases.\end{tabular} &  & \begin{tabular}[c]{@{}l@{}}Features: 697 patches/ image\\ were extracted.\\ Classifier: CNN models\\ AlexNet and GoogLeNet. Used\\ randomly selected patches of\\ 256 $\times$ 256 from large sized\\ images.\end{tabular} & \begin{tabular}[c]{@{}l@{}}Acc: 80\% and 83\% on\\ the patches using\\ AlexNet\\ and GoogLeNet.\\ Findings: GoogLeNet\\ perform better\\ than AlexNet.\end{tabular} \\ \hline
	\end{tabular}}
\end{table*}
\clearpage

\begin{itemize}
	\item Ductal carcinoma in situ. In this condition  abnormal cells takes place of  the normal epithelial
	cells which surround the breast ducts. It does not necessarily  progress towards invasive stage.
	\item Lobular carcinoma in situ. Under ths situation, the growth of abnormal cells extends to some 
	lobules of the breast and often progresses to invasive cancer.
\end{itemize}
{\textbf{ii)Invasive:}} It has two subtypes: a)Regional stage: Abnormal cells have  spread to neighbouring tissues and nearby lymph nodes (stage II or III cancers), .
b) Distant stage: A condition in which abnormal cells have metastasized to different organs or lymph nodes above collarbone (stage IIIc and stage IV cancers).

\subsubsection{\textit{Modalities of breast specimen collection}}

Patients suffering from breast cancer disease are curable if detected at an early stage. Very often symptoms
do not develop at early stages as they do develop when cancer has already reached an invasive stage
(generally correspond to stage III or IIIc and stage IV cancers ). So, a routine screening examination is
always encouraged, before symptoms actually start to divulge. This can be done by simply perceiving
a lump. Two modalities of specimen collection are normally practiced for breast cancer detection and diagnosis: a) Discharge cytology b) FNAC. FNAC is used to evaluate palpable and non-palpable breast lesions. Fine-needle or wider core needle or a surgical incision picks requisite amount of mass from several points of the site required for diagnosis. Then, microscopic analysis of breast tissue is done by an expert to determine the
extent of percolation of abnormal tissues. Nipple discharge cytology is usually performed
to very few patients who are generally asymptotic.

\subsubsection{\textit{Differential characteristics associated with cytomorphology of malignant cells}}
\begin{itemize}
	\item Large, angulated, eccentric, pleomorphic nuclei with irregular boundaries.
	\item Irregular spacing between adjacent nuclei.
	\item Background debris in large proportion.
\end{itemize}

\subsubsection{{\textit{{Significant works on automated screening of breast cell}}}}
Researchers are attempting to evolve image processing techniques \cite{Street2000Xcyt:Cancer,Niwas2010WaveletImages} since three decades to commensurate for the complex nature of breast cytological images embedded with various
degradations. We proceed to discuss the works of breast cytology
in a similar fashion: a) Segmentation based approach b) Segmentation free approach. It is worthy to
mention, preprocessing techniques such as color conversion, image normalization, contrast
enhancement etc. \cite{Saha2016Computer-aidedReview} are considered essential  prior to applying segmentation
techniques. So image enhancement techniques are often omitted during discussions without the loss of generality.\\
\textbf{\textit{i)Segmentation based approach}}

Nuclei are segmented first and then different nucleus specific features are extracted to classify benign and malignant cells. But grabbing suitable image segmentation algorithms
fit for the purpose is a challenge to the research community. Difficulty in segmentation arises mainly
due to variable structural and textural pattern of the nucleus along with various inherent noise of the
specimen. This becomes more complicated with the presence of overlapped nucleus. {In following few paragraphs, we highlight some important and popularly used segmentation techniques with the aim to develop knowledge of the existing works under each subheading.}
\\
\textit{\textbf{Contour based:}} Active contours since its introduction by Kass et al.\cite{Kass1988} in the year 1988 has been extensively studied in subsequent studies \cite{Wolberg1994MachineAspirates,Bamford1998UnsupervisedContours}. {Wolberg et al. \cite{Wolberg1993BreastAnalysis.} used manual segmentation to define nucleus boundary initially. For precise
	nucleus contour detection, they invoked snake model that can be confined to the nucleus region using proper energy function. 11 different features of isolated nuclei like size, perimeter, area, compactness,
	radial variance, concavity, texture etc. were extracted from  nucleus. Based on that, they classified the data using MSM (Multisurface Method)-Tree classifier. In their following work Wolberg et al. \cite{Wolberg1995Image-AnalysisPrognosis}, reported
	an accuracy of 97\% with a slight modification of their previous work, with maximum result obtained
	as 97.5\% for non-overlapped nuclei segmentation on their test dataset. Usage of Compact Hough
	transform for nuclei segmentation \cite{Mouroutis1998RobustModelling} and generalized hough transform with deformable models \cite{Lee2000GeneralizedTemplates} are found in literature.} Bamford et al. \cite{Bamford1998UnsupervisedContours} 
prescribed an improved
snake model to overcome initialization problem of conventional snake.
Street et al.
\cite{Street2000Xcyt:Cancer} used Hough transform to pont out circle like structures followed by
active contouring technique. They proposed an automatic diagnostic and prognosis system "Xcyt" \cite{Street2000Xcyt:Cancer} for screening of breast cancer. Hough transform  was also adopted by Hrebien et al. \cite{Hrebien2007HoughImages} followed by an automatic nuclei localization method based on (1+1) search strategy. To segment the nuclei, a combination of active contour model, watershed and grow-cut
algorithm was applied . But the technique was not fruitful for overlapping nuclei. Also, false circles were created
which was not resolved.
\\
\textit{\textbf{Texture based approach:}} Wavelet based decomposition has proved to be a powerful tool in analyzing
texture or chromatin pattern of nucleus {\cite{Weyn1998AutomatedDescription.,Jafari-Khouzani2005Rotation-invariantTransforms,Tabesh2007MultifeatureImages}}.  Weyn et al. \cite{Weyn1998AutomatedDescription.} %,
used wavelet as
chromatin pattern descriptor for semi-automated diagnosis and grading of breast tumor. However,
determination of tumor stage is hindered by increased false negative cases. Various other
works are registered using DWT as a tool \cite{Jafari-Khouzani2005Rotation-invariantTransforms, Tabesh2007MultifeatureImages} for breast cancer detection. % 
\cite{Niwas2010WaveletImages} investigated the effect of Log-Gabor wavelet filter on HSV color space. Color
wavelet features were deducted on extracted features and compared the relative performances of classifiers viz. SVM,
Naïve Bayes and ANN. Highest accuracy accorded by SVM as 98.3\% with sensitivity and specificity
of 98\% and 98.6\% respectively. Since DWT lacks phase information, complex wavelet transform
was grabbed in follow up studies \cite{10.1007/978-90-481-9794-1_50}. In \cite{ISSACNIWAS20132828}, Niwas et al.
analyzed nuclear chromatin
pattern using complex Daubechis Wavelets. Wavelet co-occurrence matrix were used to calculate
statistical features like cluster
shade and prominence, contrast, entropy, energy, local homogeneity and  maximum probability  K-NN with standard Euclidean distance
was used  to classify the images.
\\
\textit{\textbf{Region based:}}
Marker controlled watershed segmentation was studied by Yang et al.
\cite{Yang2006NucleiMicroscopy}. Hrebien et al.
\cite{Hrebien2007HoughImages} proposed   nucleus segmentation technique  using   watershed,
Active contour and Cellular automata Grow cut techniques. They reported a segmentation accuracy of 68.74\% but  taking an average of 4–5 minutes to  segment an image. Also, fake circles that were
created during the nuclei detection stage using Hough Transform  could not be removed completely. George et al.
\cite{George2014RemoteImages} suggested a fully automatic method for nuclei segmentation
of breast FNAC images. They extracted Y component of YCbCr color space for grey level conversion followed by Hough
transform to detect circular shaped structures. To eliminate the generated false circles, Otsu’s thresholding method was applied. To detect the nuclei boundaries by avoiding over-segmentation, Marker controlled watershed transform was used after that. Twelve features were extracted for
classification using MLP, PNN, LVQ, and SVM with 10 fold cross validation. 
\\
\textbf{\textit{Clustering based:}} Clustering based approach was studied in the work of \cite{Bamford1998UnsupervisedContours} using water-emersion algorithm. Seed based growing and moving k-means was propounded by Isa et al. \cite{Isa2007FineNetwork} to determine the stages of cancer. Na et al. \cite{Na2007AutomaticSegmentation} used anisotropic diffusion to determine the malignancy. Filipczuk et al. \cite{Filipczuk2013ClassifierAnalysis} approached with three level
binarization algorithm by extracting the luminance component using L=0.2126R+0.7152G+0.0722B. Initial segmentation was done using adaptive thresholding. Second level involves clustering algorithms such as k-means, fuzzy c-means (FCM) and Gaussian
mixture models (GMM)  to partition the image into nucleus, cytoplasm and
background using different color channels as features. In the final level, they combined the two
segmented images using an AND operator to give precise definition of the boundaries of the image.
But this method suffers from two major limitations. The need for determining optimal parameters and
issues associated to unsupervised clustering restricted its use to practical purpose.
\\
\textbf{Deep Learning based approach:}
Saikia et al.{\cite{saikia2019comparative} developed a CNN based deep learning  classification framework where images were augmented using techniques like cropping, shearing, rotation, mirroring, skewing  , inverting, zooming.  In subsequent phases, channel identification, histogram equalization and Otsu's thresholding were used to segment candidate nuclei . Maximum accuracy of 96.25\% was achieved using the GoogLeNet architecture .}
Kowal et al. \cite{kowal2019cell} proposed a CNN and seeded watershed based breast cytology cell segmentation model. CNN based semantic segmentation model was first applied to differentiate between the nuclei and background. After that, the  generated semantic mask was transformed into  a nuclei mask to extract the touching and overlapping nuclei. The clustered nuclei were detected by its area and roundness. Nuclei seeds were identified using conditional erosion process. The overlapping nuclei were separated by seeded watershed algorithm. With this approach, 83.4\% of benign nuclei were classified using Hausdorff distance. 
\\
\textbf{\textit{ii)Segmentation free approach}}

In this approach, instead of segmenting an image, entire image or randomly selected sub-regions are
used for feature extraction and/or classification. Based on various texture features  \cite{7232886} used Das et al. used to classify with the images. With the onset of deep learning based technique in computer vision domain especially in
medical images\cite{Litjens2017AAnalysis, HU2018134} CNN is successfully implemented to detect breast cancers \cite{Garud2017High-MagnificationNetworks,unknown}. Garud et al. \cite{Garud2017High-MagnificationNetworks} used GoogLeNet 
on randomly selected regions from the images  during training
and testing. An accuracy of 89.71\%  was reported on the test dataset after voting of classified
regions. Żejmo et al. \cite{unknown}  used  AlexNet and GoogLeNet by selecting small patches of 256 $\times$ 256 from the large sized images of 200000 $\times$ 100000
pixels. They reported accuracy of 80\% and 83\% 
on the two networks respectively. It was noticed that the accuracy observed in the CNN was still lagging behind traditional
feature based models. The accuracy can be improved by increasing  number of training samples can improve accuracy. Khan et al.
{\cite{khan2019novel} proposed a transferlearning based classification technique using VGG net, GoogLenet, ResNet. For data augmentation translation, color processing, scaling, horizontal or and/or vertical flipping, rotation and
	noise perturbation techniques were used. Features realted to circularity, compactness and roundness were extracted using CNN architectures. The classification accuracy of proposed transfer learning method was obtained as  97.525\%.}
Some notable works on breast cancer cytology are shown in Table \ref{tab:3}.
\\
\textbf{\textit{Classification:}}
\\Classifiers stand in the final lap of the image analysis system )upon which ultimate decision regarding  the nature of the specimen (whether malignant or not) is bestowed. The extent to which a classifier can
correctly classify images defines the accuracy of the system. With the finest improvement in algorithmic
complexity of the well known classifiers, researchers are able to grasp the complex and diverse nature
of images. A substantial number of scientific articles are published on a list of classifiers. Among them
Multisurface method tree (MSM-T) { \cite{Wolberg1990MultisurfaceCytology.,Street2000Xcyt:Cancer}} and KNN classifiers {\cite{Weyn1998AutomatedDescription.,kowal2011computer,ISSACNIWAS20132828}} were widely used
in the work of breast cytology image classification. Kernel induced methods like SVM, had been studied
in the works of {\cite{10.1007/978-90-481-9794-1_50,Filipczuk2013Computer-aidedBiopsies,George2014RemoteImages}} to achieve a good accuracy. Artificial neural network based approaches such as MLP \cite{Jelen2008ClassificationBiopsies} and its variant hybrid
MLP \cite{Isa2007FineNetwork}, ensemble of classifiers and Decision trees 
by Kowal et al.  \cite{kowal2011computer} were investigated. A comparative analysis of different classifiers such as SVM,
Learning Vector Quantization(LVQ), Probabilistic Neural Network, MLP was done in \cite{George2014RemoteImages} by George et al. .
\\
Studies are also motivated  towards spawning prediction of the recurrence time of the disease. As it
requires large amount of statistical data, this is usually treated as a classification problem. Logistic regression
was used by Bradley et al. \cite{Bradley2001DisparitiesSurvival} for determining recurrence time of the disease. A standard neural network trained with
backpropagation had been proposed by Street et al.  \cite{Street2000Xcyt:Cancer} to produce predictive breast cancer risk model. These prospective models are still going through several modifications to make the system  robust.

\subsection{\textbf{LUNG CANCER}}

Lungs are a pair of internal organs located on either side of chest and exchange oxygen and carbon di
oxide between the air we breathe and the blood. The inhaled air passes through main windpipe known
as trachea and conducts air into each lungs via left or right bronchus. Lungs are divided into sections
called lobes, two on the left and three on the right. The air passages divide into smaller tubes known as
tracheobronchial tree’ and connect with tiny air sacs called alveoli. The lungs are protected with a thin
tissue membrane known as pleura. Once lung cancer starts to binge to other parts of the body particularly to lymph
nodes adrenal glands, liver, brain and bones, it becomes fatal. Causes of lung cancer include smoking,
drinking or exposure to various air and water pollutants. Histological subtypes of lung cancer can
originate from different locations of the tracheobronchial tree.  
There are two major types of lung
cancer: Non-small cell lung cancer (NSCLC) which accounts for about 85\% of it and Small cell lung
cancer (SCLC) which is more aggressive than NSCLC tumors, accounts for the rest.

\subsubsection{\textit{Modalities of lung specimen collection}}

Sputum cytology \cite{Sagawa2012ScreeningFuture.} and  FNAC \cite{Mangia2015FineCancer} define changes at cellular level  to find a definite diagnosis of it and also at the same time they are not expensive.  Our present discussion is constricted only to
the cytology based diagnosis system. There are three major techniques in cytology to collect specimen
from lung nodules: i) Sputum collection ii) Bronchial Techniques iii) FNA Techniques.

i) \textit{Sputum collection:} Sputum samples are usually collected during morning for consecutive 3 days also
called ‘triple-morning test’. Sputum is collected simply when a person coughs up and so turns out to be
a minimal invasive method. The samples are then collected in a  Cytolyt  container containing 30 ml.
fixative and are sent to laboratory. As people under poverty line indulges at a greater rate in filter-less
smoking, it is recommended to undergo screening at finite intervals so they get the best out of the
economical way to test. It is extremely helpful for patients in low resource clinical settings. {It has good sensitivity for central tumors compared to peripheral tumors.}\\
ii) \textit{Bronchial Techniques:}
a) \textit{Bronchial Brushing:}  Cells are exfoliated using
a brush from the periphery of the bronchial tree. A bronchoscope
is used to guide the pathway. The cells obtained from brush are
immediately fixed in alcohol. This technique has higher sensitivity than sputum cytology . Its sensitivity and accuracy is relatively high compared to sputum cytology. because direct visualization of lesion is possible.
\\b) \textit{Bronchial Washing:} 
In this procedure, some amount of fluid is forced into the lungs through
bronchoscope and the water or the washings is retrieved back. The washing contains requisite amount
of fluid mixed with cells for cytological analysis.\\ 
c) \textit{Brochoalveolar
	Lavage:} Cells are collected from airways and alveolar lining because at the
alveolar level diseases are mostly existent. After an infusion of buffered saline solution in the alveoli
of lungs, the solution containing alveolar milieu is withdrawn back. 

\begin{table*}[h]
	\caption{{A brief overview of different works in Lung cancer cytology identification where Accuracy (Acc), Jacard Index(JI),
			Specificity (Sp), Sensitivity(Se) and Dice Similarity Coefficient (DSC) are represented in their corresponding bucket given in
			abbreviated form.}}
	\label{tab:4}
	\centering
	\resizebox{\textwidth}{!}{%
		\begin{tabular}{|l|l|l|l|l|l|}
			\hline
			\multicolumn{1}{|c|}{\textbf{Sl\#}} & \multicolumn{1}{c|}{\textbf{Author’s name}} & \multicolumn{1}{c|}{\textbf{\begin{tabular}[c]{@{}c@{}}Name/Source(N/S) and\\ Number of samples(Size)\\ in the Dataset\end{tabular}}} & \multicolumn{1}{c|}{\textbf{\begin{tabular}[c]{@{}c@{}}Method for image\\ segmentation\end{tabular}}} & \multicolumn{1}{c|}{\textbf{Features and classifier used}} & \multicolumn{1}{c|}{\textbf{\begin{tabular}[c]{@{}c@{}}Quantitative results\\ and Findings\end{tabular}}} \\ \hline
			1 & \begin{tabular}[c]{@{}l@{}} Taher et al. \\(2015)\cite{7313923}\end{tabular} & \begin{tabular}[c]{@{}l@{}}N/S: Tokyo center of lung\\ cancer in Japan.\\ Size: 100 sputum color\\ images.\end{tabular} & \begin{tabular}[c]{@{}l@{}}Mean shift\\ segmentation.\end{tabular} & \begin{tabular}[c]{@{}l@{}}Features: N/C ratio, curvature,\\ circularity,  Eigen ratio, density.\\ Classifier: ANN and SVM\end{tabular} & \begin{tabular}[c]{@{}l@{}}Acc: 97\%\\ Se: 97\%\\ Sp: 96\% \\Findings: Single sputum \\cell. Cell boundary not\\ properly delineated.\end{tabular} \\ \hline
			2 & \begin{tabular}[c]{@{}l@{}} Shajy et al.\\ (2015)\cite{Shajy2015AnalysisTransform}\end{tabular} & \begin{tabular}[c]{@{}l@{}}N/S: Regional Cancer\\ center, Thiruvanathapuram,\\ Kerala.\\ Size: 32 training and 59 testing.\end{tabular} & \begin{tabular}[c]{@{}l@{}}CLAHE and Otsu for\\ image enhancement\\ and segmentation\\ respectively.\end{tabular} & \begin{tabular}[c]{@{}l@{}}Features: DWT and GLCM separately\\ Classifiers: SVM\end{tabular} & \begin{tabular}[c]{@{}l@{}}Best result with DWT\\ Acc: 89.8\%\\ Se: 89.3\%\\ Sp: 90.3\% \\ Findings: Overlapping of cell\\ not considered. \end{tabular} \\ \hline
			3 & \begin{tabular}[c]{@{}l@{}}%Kecheril et al.\\ 2015
				Kecheril et al.\\ (2015)\cite{Kecheril2015AutomatedFeatures}\end{tabular} & \begin{tabular}[c]{@{}l@{}}N/S: Regional cancer centre,\\ Thirubanthapuram, India.\\ Size: 10 benign,24 malignant\end{tabular} & \begin{tabular}[c]{@{}l@{}}Image localization\\ using maximization\\ of determination of\\ Hessian in scale and k- \\means used for clustering.\end{tabular} & \begin{tabular}[c]{@{}l@{}}Features: 690 scale space\\ catastrophic point based on\\ feature extracted in 32 scales\\ Classifier: SVM\end{tabular} & \begin{tabular}[c]{@{}l@{}}Acc: 87.53\%\\ Se: 76.95\%\\ Sp: 92.82\% \\Findings: Only glandular \\cells are considered for\\ segmentation, while missing\\ some actual cells \\during segmentation.\end{tabular} \\ \hline
			4 & \begin{tabular}[c]{@{}l@{}}%Teramoto et al.\\ 2017
				Teramoto et al.\\ (2017)\cite{Teramoto2017AutomatedNetworks}\end{tabular} & \begin{tabular}[c]{@{}l@{}}N/S: Fujita Health University,\\ Toyoake City,Japan.\\ Size: 298 images from 76 cases.\end{tabular} & \begin{tabular}[c]{@{}l@{}}Images are resized\\ with 256 $\times$ 256 .\end{tabular} & \begin{tabular}[c]{@{}l@{}}Features and classifiers:\\ Developed 4 layer DCNN\\ with 3 convolution layers, 3\\ pooling layers, and 1 forward\\ connected layer with dropout.\end{tabular} & \begin{tabular}[c]{@{}l@{}}Acc: 71\%\\ Findings: Unsatisfactory\\ result due to non\\ customized CNN\\ architecture. \end{tabular}\\ \hline
			5 & \begin{tabular}[c]{@{}l@{}}
				Dholey et al.\\ (2018a)\cite{10.1007/978-981-10-8237-5_67}\end{tabular} & \begin{tabular}[c]{@{}l@{}}N/S: Medical College\\ Kolkata, and EKO centre\\ Kolkata, India.\\ Size: 600 images from 120 samples\end{tabular}  & \begin{tabular}[c]{@{}l@{}}GMM based hidden\\ MRF. Morphological\\ filters used to remove\\ unwanted regions.\end{tabular} & \begin{tabular}[c]{@{}l@{}}Features: SIFT, Bag of word\\ and visual dictionary after\\ clustering.\\ Classifiers: Random Forest\end{tabular} & \begin{tabular}[c]{@{}l@{}}Acc: 98.88\%\\ Se: 97.31\%\\ Sp: 99.54\\Findings: Description of\\ samples are misleading. \end{tabular} \\ \hline
			6 & \begin{tabular}[c]{@{}l@{}}
				Dholey et al.\\ (2018b)\cite{10.1007/978-981-10-7898-9_15}\end{tabular} & \begin{tabular}[c]{@{}l@{}}N/S: Medical College Kolkata, and \\EKO centre, Kolkata, India.\\ Size: 500 images from 100 samples.\end{tabular} & \begin{tabular}[c]{@{}l@{}}Random walk with K-\\ means. Watershed\\ was used to remove\\ unwanted regions.\end{tabular} & \begin{tabular}[c]{@{}l@{}}Features: 25 features.\\ (10 geometric, 14 texture\\ features and 1 color based)\\ Classifier: ANN and SVM.\end{tabular} & \begin{tabular}[c]{@{}l@{}}Using ANN (Max)\\ Acc: 97.46\%\\ Se:97.5\%\\ Sp:97.6\%\\Findings: Description of\\ samples are misleading.\end{tabular} \\ \hline \end{tabular}}
	
\end{table*} \clearpage
\begin{figure}[h]
	\centering
	\includegraphics[width=0.8\linewidth]{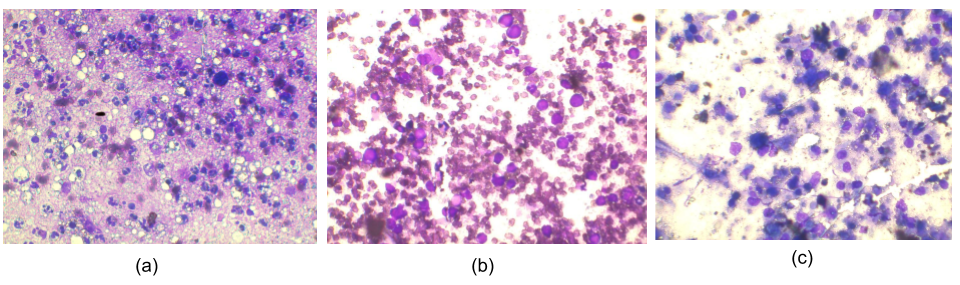}
	\caption{{Samples of cytology images of Lung (a)Benign case(MGG stain, 40X magnification)(b)Malignant case(Small cell Lung Carcinoma,MGG stain, 40X magnification),(c) (b)Malignant case(Non-Small cell Lung Carcinoma,MGG stain, 40X magnification)}}
	\label{fig:9}
\end{figure}

iii) \textit{FNA:}
Lung nodules are also aspirated using computer tomography (CT) guided FNA. It has three sub types\cite{cibas2013cytology}:{
	\\a) \textit{Transbronchial  FNA}: For the lesions located in subbronchial regions, sensitivity of TBNA is 56\% and specificity is 74\%,
	\\b) \textit{Transesophageal FNA}: It is done through endoscopy  of esophagus. It is helpful in sampling of mediastinal lymph node,
	\\c) \textit{Percutaneous FNA}: It is rapid diagnosis of pulmonary using sensitivity and specificity of 89\% and 96\% respectively. 
	
}

\subsubsection{\textit{Differential Differential characteristics associated with cytomorphology of lung malignant cells}}

\begin{itemize}
	\item Polygonal/round/fibre like cells.
	\item Abundant dense and smooth cytoplasm filled with keratin.
	\item Small hyperchromatic nucleus.
	\item Inconspicuous nucleolus.
	
\end{itemize}

\subsubsection{\textit{Significant works}}

Ample amount of works in cervical and breast cytology can be seen in the literature with varied degrees
of diversity.On the other hand, surprisingly, very few works are encountered in lung cancer cytology may be due to the lack of standard publicly available dataset.Samples of lung cytology images are shown in Fig. \ref{fig:9}.

\textbf{\textit{i) Segmentation based approach}}

\textit{\textbf{Region based:}} Kancherla et al. \cite{6595393} proposed  seeded region growing
segmentation method. On Biomada dataset they extracted 79 features related to shape, intensity, color,
wavelet based features and nucleus segmentation based features to achieve an improved recognition
accuracy of 87.8\% using bagging on Random Forest.
\\
\textit{\textbf{Color based:}} Using color information as the key discriminating factor, Taher et al.\cite{Taher2007IdentificationColor} and 
Donner et al.\cite{Donner2012CellDetection} put forward an efficient method for cell segmentation. Vineeth et al.\cite{key:article} explored the impact of color representation on classification on lung sputum images and found that Bayesian
Classifier was applied to classify the sputum cells into 3 different classes i.e.
nucleus, cytoplasm and background with misclassification error rate zero using HSV values based feature vector.
\\
\textit{\textbf{Clustering based:}}  Rachid et al.
\cite{Rachid1997SegmentationDiagnosis} proposed
an unsupervised  technique using  HNN to cluster the pixels of sputum images into nuclei, cytoplasm and background classes. To escalate the accuracy of  segmented regions  Sammouda et al. \cite{Sammouda1998SegmentationNetwork}, inserted an energy function having a cost term. However,
the method suffers from early local minimum of HNN due to which the superimposed cells could not be delineated. In the study by  \cite{10.1007/978-3-642-13681-8_60}, the sputum images were segmented into  nucleus, cytoplasm and background using HNN based module. 8-connectivity was used to find connected component
in nucleus. The proposed framework proved to be robust in
systematic  setting of the classification parameter.   \cite{Taher2012BayesianDiagnosis} proposed a
Bayesian classification framework to investigate whether a pixel in the image belongs to sputum cell or not.
HNN  segmented the nucleus, cytoplasm and background region of 88 color sputum images. The segmentation accuracies using HNN and FCM were 88.62\% and 64.91\% respectively on HSV color space. Forseberg et al. \cite{6977294} proposed an adaptive thresholding technique on whole-slide images of endobronchial ultrasound-guided TBNA. K-means clustering was used to segment the nucleus. A  processing time of 7.8 seconds/Sample was reported in the experiment. Sarvaiya et al.\cite{Sarvaiya2014DetectionSegmentation} guided a sputum image segmentation technique by classifying colors  using k-means clustering technique. Kecheril et al.\cite{Kecheril2015AutomatedFeatures} proposed a localization technique of cellular region by scale space based determinant of Hessian. Due to presence of non-cellular artifacts, Otsu’s
threshold method was adopted. K-means clustering was used to
segment nuclei. 32 scales were generated and at each scale   saddle maxima and saddle minima catastrophe points were
calculated to generate 64 dimension feature vectors.  For classification,  SVM with RBF kernel were used with 10 fold cross validation.

\textit{\textbf{Mean shift based:}}  Werghi et al.\cite{Werghi2012DetectionDetection} proposed a robust Bayesian
classifier  using RGB, HSV, YCbCr and L*a*b* colour spaces. The RGB and the
HSV showed consistency in all resolutions.
For the sputum cell segmentation, they used mean shift technique by a
judicious choice of  threshold parameter using both spatial and chromatic
information. A reasonable accuracy of 85.51\% was obtained which overule the traditional HNN showing the robustness of the used technique.\\
\textbf{\textit{ii)Segmentation free approach}}
Teramoto et al. \cite{Teramoto2017AutomatedNetworks} introduced CNN based approaches for classification of
adenocarcinoma, squamous cell carcinoma, and small cell carcinoma without any segmentation. Data augmentation techniques were used to  augment  298 samples to 5000 data. They obtained a recognition accuracy of 71.1\%. Some significant works on lung cancer cytology are shown in Table \ref{tab:4}.\\
\textbf{\textit{Classification:}}
\\
The articles on lung cytology are much confined compared to other domains of cytology. Bayesian classifiers \cite{Werghi2012DetectionDetection,Donner2012CellDetection}, ANN \cite{7313923,10.1007/978-981-10-7898-9_15}, SVM \cite{Kecheril2015AutomatedFeatures,  7313923,Shajy2015AnalysisTransform,10.1007/978-981-10-7898-9_15}, Random forest \cite{10.1007/978-981-10-8237-5_67} etc.
were popularly used in sputum cytology. \subsection{\textbf{\textit{Miscellaneous}}}

So far we have discussed different methodologies on three domains of cytology where majority  of image processing techniques are applied. {In this section,
	we will discuss the works on the rest of the fourteen types of cytology as available in literature.} We have summarized
this in a single section due to lack of sufficient works in these domains.

Some significant works are noticed in thyroid cytology {\cite{DASKALAKIS2008196,Wu2011CellRecovery,Wu2012EmbeddingData}.
Daskalakis et al. \cite{DASKALAKIS2008196} reported a discriminating technique between malignant and benign cell of	thyroid cytology image using pixel based
techniques.} From the segmented images 26 morphological and textual features were extracted and
used for classification purpose. Four different classifiers were used for comparative analysis and 
maximum accuracy of 95.7\%  was reported using majority voting based combination techniques among KNN, PNN (parametric neural network) and Bayesian network. The experiment was carried
on 53 benign and 62 malignant images, collected from the University hospital of Patras, Greece.

In  \cite{Wu2011CellRecovery,Wu2012EmbeddingData} Wu et al. used level set methods, watershed
algorithm, conditional random field, expected maximization algorithm etc. for segmentation of
multispectral cytology images works. Multispectral images were captured using photometric
sensory CCD camera.  Wu et al.\cite{Wu2012EmbeddingData}  exploited the idea of spectral data using a
multi layer {CRF} model with a bottom-up approach using local probabilistic model and an unsupervised
top-down approach with a {Probabilistic latent semantic analysis(PLSA)} representation. PLSA was used to assign a latent topic to a pixel
determined by Expectation-Maximization algorithm which was further upgraded by local mapping.
This approach showed better accuracy in segmenting nucleus than seeded watershed in color images
and also proved advantageous for low contrast and noisy environment.

Yang et al.\cite{Yang2005UnsupervisedModels} introduced an unsupervised segmentation technique for leukemia images. They estimated colour gradient and L2E into traditional GVF snakes in HSV colour space. Fifty eight lymphoproliferative cases
of four types: "chronic lymphocytic leukemia (CLL), mantle cell lymphoma (MCL) and follicle-center cell lymphoma (FCC)" were considered. The data were taken from the "Hospital of the University of Pennsylvania", Philadelphia; "Robert Wood Johnson University Hospital", New Brunswick; and "City of Hope National Medical Center", Duarte. They reported better results than unsupervised approaches such as mean shift and colour GVF snake.
\\

To identify the appropriate types of the \textit{ovarian cancer} from cytology images, Wu et al. \cite{Wu2018AutomaticNetworks} proposed
a CNN based approach using AlexNet. 85 labelled specimens in 4 categories namely:  "serous carcinoma, mucinous
carcinoma, endometrioid carcinoma, and clear cell carcinoma"  were collected from
"Xinjiang Medical University". 1848 ovarian cancer cytological images of
1360$\times$1024 pixels dimensions were considered. They used rotation based augmentation technique %Please add the following required packages to your document preamble:

to increase the dataset into  20,328 samples. The images were divided into four sub images and resized them
to 227$\times$227. After data augmentation they got
accuracy of 78.20\% using AlexNet on 10 fold cross validation.

% \usepackage{graphicx}
% \usepackage[normalem]{ulem}
\begin{table*}[]
	\caption{A brief overview of different works in other cytology identification where Accuracy (Acc), Jacard Index(JI), Specificity (Sp), Sensitivity(Se)and Dice Similarity Coefficient (DSC) are represented in their corresponding bucket given in abbreviated form. }
	\label{tab:my-table}
	\resizebox{\textwidth}{!}{%
		\begin{tabular}{|c|c|c|c|c|l|}
			\hline
			\textbf{SL\#} & \textbf{Author}       & \textbf{\begin{tabular}[c]{@{}c@{}}Name/ Source\\ and number of\\ samples(Size) in dataset\end{tabular}}                                                                         & \textbf{Method in Image Segmentation}                                                                                                                                                                                                                                            & \textbf{Features and Classifiers used}                                                                                                                                                                              & \multicolumn{1}{c|}{\textbf{Quantitative results and Findings}}                                                                                                                                        \\ \hline
			1.            & {Wu et al.(2009)\cite{wu2009multispectral}}      & Thyroid FNAC smears                                                                                                                                                              & \begin{tabular}[c]{@{}c@{}}Morphological self dual reconstraction\\ using Watershed algorithm.\end{tabular}                                                                                                                                                                      & Bhattacharyya distance                                                                                                                                                                                              & \begin{tabular}[c]{@{}l@{}}Acc.- 95.19\%\\ FP- 2.12\%\\ Findings:Only segmentation of the images.\\ No standart dataset\end{tabular}                                                                   \\ \hline
			2.            & {Gopinath et al.(2013)\cite{gopinath2013computer}} & \begin{tabular}[c]{@{}c@{}}Thyroid FNAB collected from\\ on-line image atlas of Papanicolaou\\ Society of Cytopathology\end{tabular}                                             & \begin{tabular}[c]{@{}c@{}}Morphological operations\\ Watershed transformation\end{tabular}                                                                                                                                                                                      & \begin{tabular}[c]{@{}c@{}}Features:\\ Two-level discrete wavelet decomposition\\ GLCM, Gabor filters\\ Classifiers: k-NN, ENN(Elman neural network),SVM\end{tabular}                                               & \begin{tabular}[c]{@{}l@{}}Acc: 93.33 \% in\\ ENN classifier by training statistical features\\ Findings: CAD based system\\ for multi-stain images.\end{tabular}                                      \\ \hline
			3.            & {Sanyal et al.(2018) \cite{sanyal2018artificial}}   & \begin{tabular}[c]{@{}c@{}}Thyroid FNAC smears\\ collected from tertiary\\ care centers of North\\ India\end{tabular}                                                            & Not applicable                                                                                                                                                                                                                                                                   & \begin{tabular}[c]{@{}c@{}}ANN\\ (Artificial Neural Network)\end{tabular}                                                                                                                                           & \begin{tabular}[c]{@{}l@{}}Combined result in 10x and 40x magnification:\\ Se: :90.48\%\\ Sp: 83.33\%\\ negative predictive value :96.49\%\\ Acc: 85.06\%\\ Findings: No standrad dataset\end{tabular} \\ \hline
			4.            & {Dimauro et al.(2018)\cite{nasal}}  & \begin{tabular}[c]{@{}c@{}}superficial cells from the nasal\\ mucosa collected from Rhinology\\ Clinic of the Otolaryngology\\ Department of the University of Bari\end{tabular} & \begin{tabular}[c]{@{}c@{}}Otsu thresolding algorithm,morphological\\ opening operation follwed by labelling,\\ marking objects\end{tabular}                                                                                                                                     & \begin{tabular}[c]{@{}c@{}}Classifiers: CNN(ConvNet)\\ 1st approach:\\ Data augmentation: geometric transformations\\ (reflection,rotation, translation)\\ 2nd approach:\\ hyperparameter-optimization\end{tabular} & \begin{tabular}[c]{@{}l@{}}Result:\\ Acc: 94.0\%\\ Se: 96.4\%\\ Sp: 82.5\%\\ Findings: Low cost system for preparing\\ report of rhino-cytogram.\end{tabular}                                          \\ \hline
			5.            & {Hossain et al.(2019)\cite{hossain2019renal}}  & \begin{tabular}[c]{@{}c@{}}Renal cytology collected from\\ National Cancer Institute, USA\end{tabular}                                                                           & \begin{tabular}[c]{@{}c@{}}K-means clustering to classify\\ background and nucleus region.\\ SVM to differentiate between\\ normal and abnormal nuclei regions.\\ Selective search algorithm is used to\\ detect irregular shaped structures in\\ abnormal regions.\end{tabular} & \begin{tabular}[c]{@{}c@{}}Features: patches of normal and abnormal cells.\\ Classifier: RCNN used for normal and\\ abnormal cell detection\end{tabular}                                                            & \begin{tabular}[c]{@{}l@{}}Precision:99.01\%\\ Recall:98.7\%\\ F-measure:98.8\%\\ Findings:proliferation rate estimation for\\ successful prognosis of the disease.\end{tabular}                       \\ \hline
			6.            & {Dimauro et al.(2019)\cite{dimauro2019nasal}}  & \begin{tabular}[c]{@{}c@{}}14 Nasal cytology images\\ collected from Rhinology Clinic\\ of the Otolaryngology Department\\ of the University of Bari\end{tabular}                & \begin{tabular}[c]{@{}c@{}}K-means clustering to classify\\ background and nucleus region.\\ SVM to differentiate between\\ normal and abnormal nuclei regions.\\ Selective search algorithm is used to\\ detect irregular shaped structures in\\ abnormal regions.\end{tabular} & Classifier: CNN                                                                                                                                                                                                     & \begin{tabular}[c]{@{}l@{}}Result of test set on 7 class\\ problem\\ Se: 98.6\%\\ Sp: 99.7\%\\ Acc: 99.4\%\\ Findings: Manual acquisition of fields.\\ No standrad dataset.\end{tabular}               \\ \hline
		\end{tabular}%
	}
\end{table*}

\section{COMMERCIALLY AVAILABLE SYSTEMS}

The core idea to insert robust features with proper segmentation techniques is to build smart  user-end systems
capable of handling large amount of data. Thus, it could reduce the
workload of the cytotechnologists. Evolution of automation system for cytological images has started
since 1950. With the improvement in slide preparation techniques, some systems became obsolete and
some modified accordingly. New systems evolved with better techniques and high computation powers.
There are a series of systems available for cervical cytology, but for breast cancer and lung cancer very
few systems are found at one's disposal. Unfortunately most of the systems are not commercially available in the
market. Some of the popular commercially available systems are recorded in the Table \ref{tab:5}.

\begin{table*}[h]
	\caption{\label{tab1}Different commercially successful Cytology based systems}
	\label{tab:5}
	\centering
	\resizebox{\textwidth}{!}{%
		\begin{tabular}{|l|l|l|l|}
			\hline
			\multicolumn{1}{|c|}{\textbf{\begin{tabular}[c]{@{}c@{}}Manufacturer and\\ year\\ of\\ manufacturing\end{tabular}}} & \multicolumn{1}{c|}{\textbf{\begin{tabular}[c]{@{}c@{}}Cancer\\ pre-screening\\ systems\end{tabular}}} & \multicolumn{1}{c|}{\textbf{Description}} & \multicolumn{1}{c|}{\textbf{{Assessment}}} \\ \hline
			\begin{tabular}[c]{@{}l@{}}Airborne Instruments\\ Laboratory, Inc.,\\ N. Y., 1950\end{tabular} & Cytoanalyzer & \begin{tabular}[c]{@{}l@{}}An electronic optical machine to detect abnormal pap smear \\cells. The machine consists of a scanning  microscope, computer\\ and analyzer, and recorder. The scanner examines the significant\\ area of the smear and converts the optical information into an\\ electric beam which is passed to the computer and analyzer.\end{tabular} & \begin{tabular}[c]{@{}l@{}}Two sets of experiments\\ were conducted between\\ 1958-1960 and results\\ reveal high false negative\\ cases which hindered to\\ make it commercially available\\ \cite{Tolles1955TheCE,shapiro2004evolution}.\end{tabular} \\ \hline
			\begin{tabular}[c]{@{}l@{}}Watanabe and Toshiba,\\ 1972\end{tabular} & CYBEST & \begin{tabular}[c]{@{}l@{}}The pap smear screening system that uses object extraction \\techniques like thresholding and differential  approaches. It uses\\ morphological features nuclear size, N/C ratio, optical density,\\ nuclear shape, chromatin pattern for image analysis. The model\\ was upgraded to  model 2 (1974), model 3 (1978) and model 4 (1981)\\ accordingly with improved designs and reduced turnaround time.\end{tabular} & \begin{tabular}[c]{@{}l@{}}Field tests of CYBEST model 4\\ shows occurrences of greater\\ false positive cases. CYBEST \\model 4 takes nearly 3 minutes per\\ specimen to evaluate\\ \cite{tanaka1987automated}.\end{tabular} \\ \hline
			DJ Zahniser ,1979 & \begin{tabular}[c]{@{}l@{}}BioPEPR system \\ %\cite{Zahniser1979BioPEPR:Smears}
			\end{tabular} & \begin{tabular}[c]{@{}l@{}}It was designed to prescreen cervical smear based on cellular \\morphology nuclear area, nuclear optical density, nuclear texture,\\ and N/C ratio. Additional programs were designed to recognize\\ artifacts, overlapping nuclei  and leukocytes.\end{tabular} & \begin{tabular}[c]{@{}l@{}}A high false alarm rate of 24\%,\\ restricted commercial feasibility\\ and also efficiency was highly\\ dependent on quality of smears.\\ Analysis rate was 4\\ minutes/smear \cite{haralick2012pictorial}.\end{tabular} \\ \hline
			\begin{tabular}[c]{@{}l@{}}BD Diagnostics\\ (Sparks, MD) \end{tabular} & \begin{tabular}[c]{@{}l@{}}Focal Point Slide Profiler\\ (FDA approval in 2006)\end{tabular} & \begin{tabular}[c]{@{}l}It is an FDA approved system that screens the entire slide.\\ A Focal Point Score was derived from the slide using a model\\ based on different features such as nuclear size, contour,\\ ratio with cytoplasm, integrated optical density.\\ A score less than a threshold is separated into \\‘’Further Review’’ category\end{tabular} & \begin{tabular}[c]{@{}l@{}}For glandular abnormalities, FocalPoint\\ screened slides need to be reviewed \\exhaustively regardless of the quintile \\ranking. Also it was not cost\\ effective and required heavy \\technical maintenance\\ \cite{pantanowitz2014practical}.\end{tabular} \\ \hline
			\begin{tabular}[c]{@{}l@{}}Tripath (1998)\\ formed with three\\ companies Neopath,,\\ Neuromedical and\\ Autocyte\end{tabular} & \begin{tabular}[c]{@{}l@{}}Autoprep 300\\ (FDA approved\\ in 1998)\end{tabular} & \begin{tabular}[c]{@{}l@{}}A first FDA approved rescreening system that infused neural\\ network using morphological features into low level programming.\\ It used two resolution levels: low resolution to map the specimen\\ and high resolution for selecting ROI. Only suspicious samples \\detected was labeled as ‘Required Visual Inspection'. The image\\ processing system was developed using specific system board ASIC.\end{tabular} & \begin{tabular}[c]{@{}l@{}}The system is confirmed\\ for 25\% cases without\\ further screening as\\ normal cases. Rest of the\\ cases are categorized into\\ five different stages of\\ abnormality\\ \cite{valente2001cytology}.\end{tabular} \\ \hline
			\begin{tabular}[c]{@{}l@{}}Hologic, Inc., taken over \\ from Cytyc Corp,\\ Marlborough, MA\end{tabular} & \begin{tabular}[c]{@{}l@{}}ThinPrep® Integrated\\ Imager (T- 3000)\\ FDA approval\\ in 2018.\end{tabular} & \begin{tabular}[c]{@{}l@{}}Designed for pap tests, it consists of a single desktop system\\ using an imaging station and microscope and directs \\technicians to examine only potentially abnormal areas.\end{tabular} & \begin{tabular}[c]{@{}l@{}} Increased throughput by detecting\\ only suspicious cells without\\ complete manual slide review. \\It reviews a slide in $\sim$90 seconds \\ \cite{manual}.\end{tabular} \\ \hline
			\begin{tabular}[c]{@{}l@{}}Neuromedical\\ Sciences Inc (NSI)\\ Late 1990s\end{tabular} & \begin{tabular}[c]{@{}l@{}}PAPNET Systems\end{tabular} & \begin{tabular}[c]{@{}l@{}}A rescreening system that infused neural network into low level\\ programming. It used two resolution levels: low resolution to\\ map the specimen and high resolution for selecting ROI. Only\\ suspicious samples detected was labeled as ‘Required Visual \\Inspection’.\end{tabular} & \begin{tabular}[c]{@{}l@{}}Not cost effective and no\\ hint of malignancy either.\\ In the end of 1999, NSI\\ was exhausted of capitals and \\ declared an economic\\ failure to run the project\\ \cite{koss1994evaluation}.\end{tabular} \\ \hline
			\begin{tabular}[c]{@{}l@{}}C-DAC along with\\ the RCC,\\ Trivandrum\end{tabular} & CerviSCAN & \begin{tabular}[c]{@{}l@{}}This pap smear screening system has a Piezo server controller \\connected to a micrposcope. It captures the information of a\\ nucleus at various foci and stacks the information for every\\ nuclei.\end{tabular} & \begin{tabular}[c]{@{}l@{}}Creating image stack required\\ expert intervention and huge\\ memory. Also, collecting\\ information of nucleus at\\ various focus hindered by\\ background debris \\ \cite{tucker1976cerviscan}.\end{tabular} \\ \hline
			\begin{tabular}[c]{@{}l@{}}W.N Street,1990,\\ University of\\ Wisconsin Madi-\\ sonin.\end{tabular} & “Xcyt project’’ & \begin{tabular}[c]{@{}l@{}}A breast cancer diagnosis and prognosis system.\end{tabular} & More false negative cases \cite{street2000xcyt}. \\ \hline
			\begin{tabular}[c]{@{}l@{}}Roger A. Kemp\\ 2007\end{tabular} & LungSign test & \begin{tabular}[c]{@{}l@{}}LungSign is a fully automated system for analysis of sputum\\ specimens to scan slides automatically. Using cellular\\ morphology it produces a score for individual specimen.\end{tabular} & \begin{tabular}[c]{@{}l@{}}LungSign is an effective tool \\for detecting stage 1 cancer \\and needs to be upgraded\\ for high stage cancers \\ \cite{kemp2007detection}.\end{tabular} \\ \hline
		\end{tabular}%
	}
\end{table*}\clearpage
\section{AUTHORS VIEW}

{During the journey 
	towards} automation in cytology, we  clearly make out the goals and
achieved solutions. The primary goal is to help cytotechnologists to reduce turnaround time with lesser
“false” cases. Though none of the false cases are desired either, false negative cases should be totally
delimited, from diagnostic point of view. Because one false negative case would leave the patient
untreated leading to a fatal situation. False positive cases, on the other hand, puts the patient in a
traumatic condition. Thus, machine driven output should be at least comparable to human evaluation.
The second goal is obviously the cost effectiveness, so that it can be run in low resource
clinical settings.

\begin{figure}[h]
	\centering
	\includegraphics[width=0.8 \linewidth]{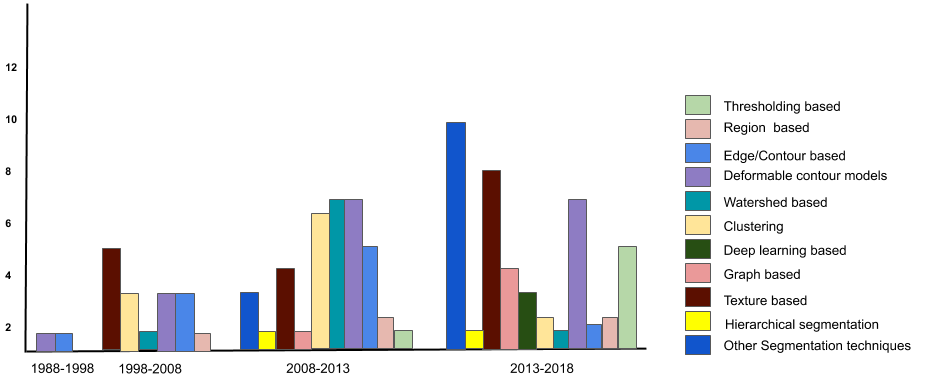}
	\caption{{Major segmentation techniques and their usages in last three decades on cytology images}}
	\label{fig:10}
\end{figure}

Some of the major segmentation techniques found extensively in literature are jotted down in Table \ref{tab:6}. It is observed that a lot
of experiments have been done on cervical and breast cancer detection whereas the number of
algorithms explored for other types of cancers are very few. This survey points out the progress of
automation in terms of research to reach to a viable commercial output. During the period of 1988
to1998, though meagre amount of cytology based research works are registered, most of the researches
tend to concentrate towards contour and deformable contour models as depicted in Fig. \ref{fig:10}. During the period of 1998 to 2008
texture based features using wavelet analysis were in vogue. During the last decade, deformable contour
models and clustering based segmentation approach showed unprecedented pace jointly bagging almost
45\% of the research works using rest of the segmentation techniques. As neural network based deep
learning techniques has become the cutting edge technology from past few years due to its automatic feature learning mechanism, it is slowly replacing
previously used classification techniques using KNN, Random Forest, SVM etc. Despite its successful implementation in several areas of research  its performance is largely dictated by a exhaustive dataset which has paucity in cytology domain .

Though there are diverse and plenty of works in cervical and breast cytology domain, a little less is
realized in lung cytology. {At the same time the research works on histopathology images of lung are much more pronounced and diverse compared to cytology.} It is observed, that methodologies in lung cytology tend to concentrate
towards particular techniques. So, techniques also lack diversity at the end. Also, majority of the
techniques produce only segmented images. Thus, end user’s interference is mandatory. Again, works
exist only in sputum cytology whereas other modalities of cytology in lung cancer detection are almost
{unattended creating a large void and opportunity to work on those modalities, thereby exploring possibilities to generate high degree of recognition accuracy.}

Developing certified systems for automotive application requires a deep insight of the functionality of
the design including power, cost and time to market to help ensure market success. Although, some
automated devices have become de facto medical standard in few parts of the world, third world
countries like, India, are still to exploit the advantage of the screening devices on a full scale. So, efforts
are now streamlined towards producing screening unit that can be operated in a
semi-automated fashion. Researches are presently heading towards producing intelligent and remote
web-based diagnostic system which can handle a large group of patients with reduced false cases. It is
observed that false cases are almost part and parcel of automated systems. Despite several
disadvantages, many software and hardware automated systems are available in laboratories that act as human companion, assisting in various decision making processes by generating a reference from a machine generated output. Existing systems though require end
users interaction, nevertheless they reduced the workload to a greater extent. Thus, cytotechnologists  apart from reviewing slides can also join in
research activities. 

\begin{table*}[h]\caption{\label{tab1}Major segmentation algorithms used in cytology images}
	\label{tab:6}
	\centering
	\resizebox{\textwidth}{!}{%
		\begin{tabular}{|l|l|l|}
			\hline
			\multicolumn{1}{|c|}{\textbf{Sl\#}} & \multicolumn{1}{c|}{\textbf{Segmentation techniques}} & \multicolumn{1}{c|}{\textbf{ Methodologies}} \\ \hline
			1 & Thresholding based & \begin{tabular}[c]{@{}l@{}}a. Otsu thresholding \cite{7371285}, b. Modified Otsu with class prior probability \cite{Kumar2016AnSegmentation}\\ c. Adaptive thresholding method \cite{Zhang2011ACytology};\cite{Agarwal2015Mean-shiftImages};\cite{Lee2016SegmentationRefinement}, d. Iterative adaptive classified algorithm \cite{Zhao2016AutomaticMRF}\end{tabular} \\ \hline
			2 & Region based & \begin{tabular}[c]{@{}l@{}}a. Seed based region growing \cite{6595393}, b. MSBRG \cite{Mat-Isa2005AutomatedAlgorithm}; \cite{Lin2009DetectionNucleus}, c. MSER \cite{10.1007/978-3-642-40811-3_57} \\d. Grow-Cut \cite{article1}, e. Anisotropic diffusion and  anisotropic kernel mean shift \cite{7371285}\end{tabular} \\ \hline
			3 & Edge/Contour based & \begin{tabular}[c]{@{}l@{}}a. CNC detector\cite{Tsai:2008:NCC:1371261.1371379}; \cite{Pai2012NucleusImage}, b. EENCC detector\cite{Yang-Mao2008EdgeImages}, c. Sobel edge detector\cite{sobel}\\ d. Non-maximum suppression \cite{Lin2009DetectionNucleus}, e. Hough transform\cite{Street2000Xcyt:Cancer}; \cite{Hrebien2007HoughImages}, f. Compact Hough transform \cite{Mouroutis1997CompactDetection}\\ g. Two-group edge enhance method \cite{Lin2009DetectionNucleus}, h. Laplacian, Prewitt, Roberts, Robinson\cite{Mat-Isa2005AutomatedAlgorithm},\\ i. Superpixel Partitioning and Cell-Wise Contour Refinement \cite{Lee2016SegmentationRefinement}\end{tabular} \\ \hline
			4 & Deformable contour models & \begin{tabular}[c]{@{}l@{}}a. Active Contour Model \cite{Yang2005UnsupervisedModels};\cite{7371285}, b. Hough transform with deformable models \cite{Lee2000GeneralizedTemplates},\\ c.  Adaptive active contour modelling \cite{Zeng2016UnsupervisedModelling}, d.  Snake model \cite{Wolberg1993BreastAnalysis.};\cite{Bamford1998UnsupervisedContours};\cite{Niwas2010WaveletImages}, e. GVF snakes\cite{Malek:2009:ABC:1527246.1527265},\\ f. Radiating GVF Snake \cite{Li2012CytoplasmSnake};\cite{Sajeena2015CervicalCells}, g. Dynamic sparse contour and GVF Snake \cite{Guan2015AccurateSO}, h. Viterbi search-based dual ACM \cite{Ramli2004DiagnosisNetwork},\\ i. Level set method\cite{Mustafa2007ColourImages};\cite{Lin2009DetectionNucleus};\cite{Wu2011CellRecovery}, j. Joint level set \cite{10.1007/978-3-642-40811-3_57}, k. Multiple level set\cite{Lu2015AnCells},\\ l. Level set method active contour model\cite{6485273}\\ m. Multi-step level set method \cite{Islam2015Multi-stepCells}\end{tabular} \\ \hline
			5 & Watershed based & \begin{tabular}[c]{@{}l@{}}a. Watershed based \cite{article1}; \cite{Plissiti2011CombiningImages};\cite{Muhimmah2012AutomatedTransformation};\cite{Orozco-Monteagudo2013Pap-smearWatershed}; \cite{10.1007/978-981-10-5780-9_15}, b. Multi-pass fast watershed\cite{10.1007/978-981-10-7898-9_15},\\ c. Multi scale Watershed\cite{Kale2010SegmentationImages}, d. Colour based watershed \cite{Lezoray2002HistogramImages}, e. Hierarchical watershed \cite{Orozco-Monteagudo2012CombinedExtraction}\end{tabular} \\ \hline 6 & Texture based & \begin{tabular}[c]{@{}l@{}}a. Coarseness \cite{Zahniser1979BioPEPR:Smears}, b. Texture Filter Bank \cite{Bora2017AutomatedDysplasia}, c. Scale space features\cite{Kecheril2015AutomatedFeatures}\\ d. Wavelet transforms \cite{Weyn1998AutomatedDescription.}; \cite{ISSACNIWAS20132828}; \cite{AnnaLakshmi2017AutomatedICM}, e. Gray Level Co-Occurrence Matrix \cite{Walker1997AdaptiveCytology}; \cite{Sukumar2016ComputerClassifier}\\ f. Gabor\cite{Niwas2010WaveletImages};\cite{Rahmadwati2011CervicalFilters};\cite{Muhimmah2012AutomatedTransformation};\cite{Filipczuk2013ClassifierAnalysis}, g. Rotation invariant LBP \cite{Ojala2002MultiresolutionPatterns}\\ h. Conventional LBP\cite{Filipczuk2013ClassifierAnalysis}, rotation invariant patterns, local patterns with anisotropic\\ structure, completed local binary pattern (CLBP) and local ternary pattern (LTP)\cite{Guo2012DiscriminativeDescription}\end{tabular}
			\\ \hline
			7 & Graph based & \begin{tabular}[c]{@{}l@{}}a. Graph cut \cite{Zhang2014AutomatedCuts}, b. Global and local graph cuts\cite{ZHANG2014369};\cite{Wu2018AutomaticNetworks}\\ c. Graph search \cite{Zhang2017Graph-basedCytology}, d. Graph-search based MRF- \cite{Zhao2016AutomaticMRF}\end{tabular} \\ \hline 8 & Clustering & \begin{tabular}[c]{@{}l@{}}a. Spatial k-means \cite{Li2012CytoplasmSnake}, b. k-means clustering algorithm\cite{Mat-Isa2005AutomatedAlgorithm};\cite{6977294}\\ c. Fuzzy C means clustering\cite{Kim2007NucleusAlgorithm};\cite{Sharma2016AnBPNN};\cite{Saha2016SpatialImages};\cite{Indrabayu2017AClustering} %;\cite{Filipczuk2013ClassifierAnalysis}
				, d. Spatial patch based Fuzzy C-means clustering \cite{Chankong2014AutomaticFilter},\\ e. GMM based \cite{1647674};\cite{kowal2011computer};\cite{7789663}, f. GMM with HMRF\cite{10.1007/978-981-10-8237-5_67}, g. Superpixel based MRF \cite{Zhao2016AutomaticMRF},\\ h. SLIC\cite{Plissiti2015SegmentationSuperpixels}, i. Superpixel Partitioning and Cell-Wise Contour Refinement \cite{Lee2016SegmentationRefinement},\\ j. Entropy based superpixel method \cite{Mitra2018IdentificationApproach}, k. SuperPixel with Voronoi\cite{Ushizima2014SegmentationOS}\end{tabular} \\ \hline
			9 & Hierarchical segmentation & \begin{tabular}[c]{@{}l@{}}a. Hierarchical shape approximation, and shape regularization \cite{7371285}\\ b. Spectral and shape information based non-parametric\\ hierarchical segmentation algorithm\cite{7789663}\end{tabular} \\ \hline
			10 & Deep learning based & \begin{tabular}[c]{@{}l@{}}a. Multiscale CNNS \cite{7562400}, b. Multiscale CNN and graph based partitioning \cite{Song2015AccuratePartitioning}\\ c. Deep learning and dynamic shape modeling\cite{TAREEF201728}\end{tabular} \\ \hline
			11 & \begin{tabular}[c]{@{}l@{}}Other Segmentation\\ techniques\end{tabular} & \begin{tabular}[c]{@{}l@{}}a. Colour based \cite{Zhang2011ACytology};\cite{Donner2012CellDetection};\cite{Boughzala2016AutomaticSO}; \cite{Indrabayu2017AClustering};; \cite{Carneiro2017AEffusions}, b. Phansalkar’s local search\cite{Ushizima2014SegmentationOS}\\ c. Intersecting Cortical Model (ICM) \cite{7482168};\cite{AnnaLakshmi2017AutomatedICM}, d. Spatially adaptive active physical model \cite{Plissiti2012OverlappingModel}\\ e. Minimax optimization of an energy functional\cite{Agarwal2015Mean-shiftImages}, f. Mean Shift\cite{Wu2018AutomaticNetworks}\\ g. QPSO with Fuzzy KNN\cite{Zhang2017Graph-basedCytology}, h. Star shape prior and Voronoi energy term\cite{7163846}\end{tabular} 
			\\ \hline
		\end{tabular}%
	}
\end{table*}

\section{CONCLUSION}

This article reviews and captures the diversity of the state-of-the-art methodologies in the analysis of
cytology images. It starts with a brief introduction of cytology. Steps required to prepare a cytology specimen
are subsequently discussed. In the following section we tabularize 17 different types of cytology along with the modalities of specimen collection from patients and also their challenges in detecting the disease. We
have pointed out that most of the existing literature revolve around cervical and breast cytology, and
other domains in cytology do not receive much of researcher’s attention. We have also realized that
some segmentation frameworks surpass all the previous results following  the recent trends of Deep
learning module, while other traditional/conventional methodologies pay attention to segment the cells using handcrafted features of nucleus. To be precise, we can now encapsulate some of the
outcomes of this article as possible future directions to bring the cytology commercial systems into the
mainstream. Below are summarized some of the areas in cytology where there are ample scope for upgradations:
\begin{itemize}
	\item Lack of freely available standard datasets for cytological images, except for pap smear
	cytology. Again, the size of the available datasets are not sufficient enough to train a deep learning module effectively.
	
	\item Appropriate segmentation algorithm is yet to be designed for segmentation of clusters
	of overlapped nuclei. In addition, majority of the segmentation algorithms are developed
	focusing on supervised data, whereas robust unsupervised algorithms need to be developed for
	automatic labelling of data.
	\item Absence of standardized staining technique hinders the segmentation techniques to be
	appropriately implemented.
	\item Processing high resolution images with conventional segmentation techniques is time taking.
	\item Non availability of viable commercial systems apart from pap smear images.
	\item There exist no single system for handling cytological images in different domains.
\end{itemize}
Possible future directions to successfully run the existing systems can be enumerated below:
\begin{itemize}
	\item  The limitation of available standardized dataset can be overcome by synthetically
	generating realistic dataset using contemporary deep learning techniques, such as Adversial
	Neural Network, Variational Autoencoder etc.
	\item Appropriate resizing techniques \cite{ghosh2019reshaping} need to be grabbed for feeding deep learning modules
	\item Parallelization of algorithms can aid to make computational time faster.
	\item Implementation of techniques for hand–held devices using different embedded modules, such
	as Raspberry Pi, Arduino etc.
\end{itemize}

The automatic understanding the nature of cytology images is a challenge to the
researchers because of their diverse nature and presence of unusual artifacts. Some systems that are
available for cervical, breast  and lung cancer are not robust enough to deal with all kinds
of the data existing globally, so could not  be marketed in a large scale. Also, production of high precision screening machines are not very cost-effective. All these factors coagulated to hinder
the current screening systems reaching the third world countries like India. So, efforts are now
streamlined towards finding a feasible solution or a screening unit that can at least be run in a
semiautomatic fashion. The success in this regard, till date, is limited. Still there is a lot of scope to
work in this domain due improvements in data collection methodology and staining techniques.
We, finally, hope that this survey will help the researchers to comprehend the latest state-of-the-art
methods and progress of cytology based research for systematic and selective design of algorithms and
test ideas to develop a conceptual framework suitable for analysis in relevant problem domain.

\section*{Acknowledgement}
This work is partially supported by SERB (DST), Govt. of India  sponsored project ( order no. EEQ/2018/000963 dated 22/03/2019). Authors are also thankful to the members of "Theism Medical Diagnostics Centre", Kolkata, India and "Saroj Gupta Cancer Centre \& Research Institute", Thakurpukur, Kolkata, India.

\section*{Supplementary Information}
{\textbf{Specimen Collection}
}

\begin{figure*}[h]
	\centering
	\includegraphics[width=0.8\linewidth]{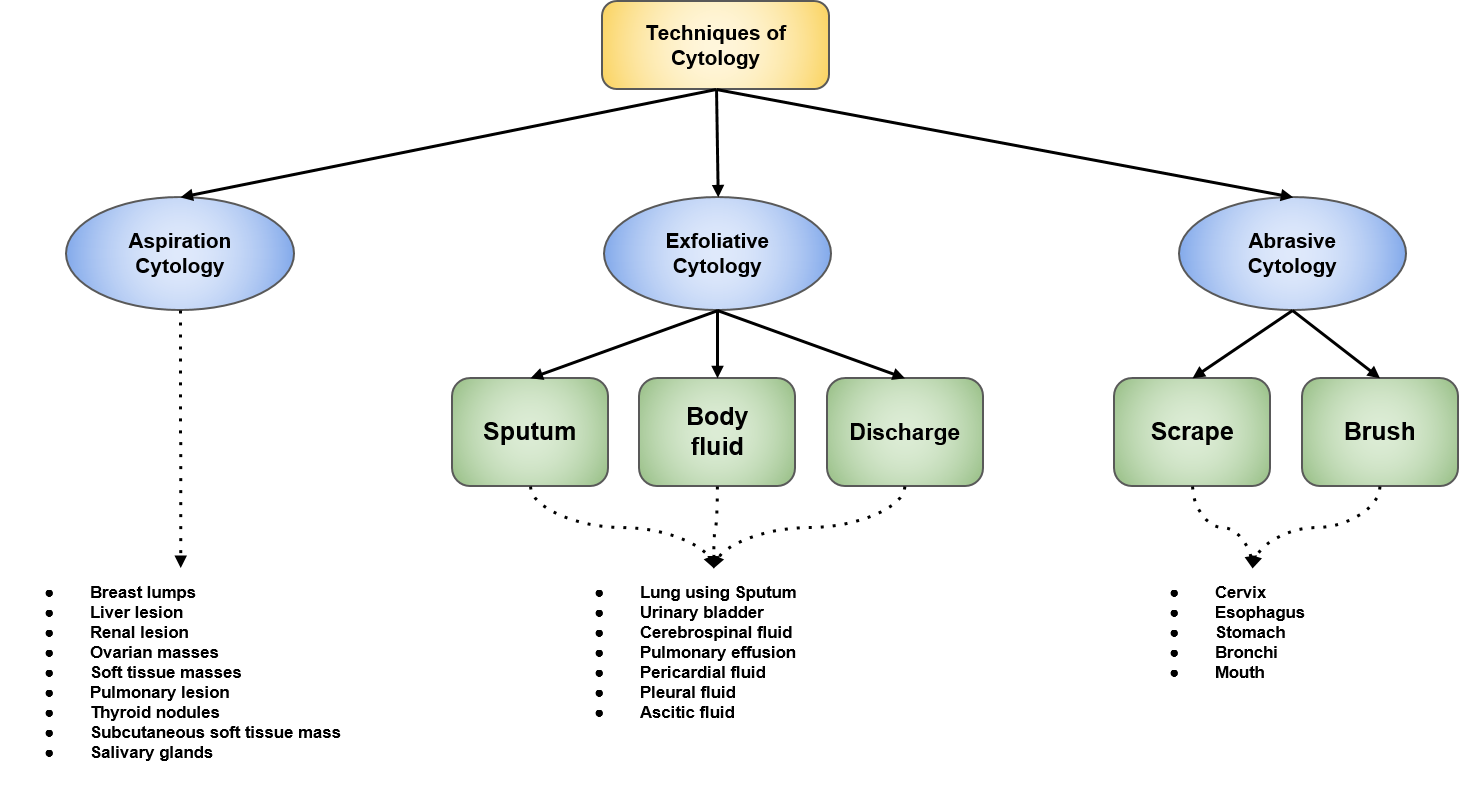}
	\caption{{Different types of specimen collection techniques and their associated domains in cytology}}
	\label{fig:2}
\end{figure*}

In any cytological test, collection of specimen is the first step where pathologists collect cells from the
predetermined mass for pathological analysis. Based on collection techniques cytology can be broadly categorized under three heads: 1)\textit{ Aspiration cytology} 2) \textit{Exfoliative cytology} 3) \textit{Abrasive cytology} as shown in Fig \ref{fig:2}.

\begin{enumerate}[]
	\item{\textbf{Aspiration cytology}}
	In this study or procedure, cells are aspirated \textit{i.e} taken out by using a fine needle normally of 23-25
	gauge from body fluid, cyst or  palpable mass of clinical suspicion. Prior to injection, the suspected region
	is cleaned by swabbing with a cotton soaked in antiseptic solution. A well known example of aspiration cytology is FNA. It is extensively used in the detection and
	diagnosis of breast lumps, liver lesions, renal lesions, ovarian masses, soft tissue masses, pulmonary
	lesions, thyroid nodules, subcutaneous soft tissue mass, salivary gland, lymph nodes etc. \cite{LopesCardozo1980TheLymphomas.}. This process is relatively faster, safer, cheaper, non-invasive and less
	painful method when compared to surgical biopsy \cite{Domanski2007Fine-needleChallenges}. Serious
	complications are very rare and include redness, soreness and minor hemorrhage. First FNAC
	was done at Maimonides Medical Center, United States, in 1981 successfully which paved
	the way to an almost painless and trauma free diagnosis process.
	
	\item{\textbf{Exfoliative Cytology}}
	It is the study of micro examination of desquamated cells from the body surfaces or cells that are
	harvested by rubbing or brushing a lesional tissue surface. It consists of three sub types:-

	(i) Body fluid cytology: This includes:
	\begin{itemize}
		\item Urine
		\item Cerebrospinal fluid (CSF) is a fluid that surrounds the brain and spinal cord.
		\item Pleural fluid (pulmonary effusion) is an accretion of fluid in the lining of tissues between lungs and the chest cavity.
		\item Pericardial fluid is secreted by the serous layer of the pericardium into the pericardial cavity surrounding the heart.
		\item Ascetic fluid also called ascites or peritoneal fluid, refers to abnormal accumulation fluid in peritoneal or abdominal cavity.
	\end{itemize}
	(ii) Discharge cytology: In this category, cytological examination of breast secretions are carried out to diagnose the disease.
	(iii) Sputum cytology: Sputum is a popular exfoliative cytology  which is  usually used
	for lung disease detection. Sputum (phlegm) is a mixture of mucus and saliva consisting of exfoliated epithelial cells that line the respiratory tract. It is spontaneous (often aerosol induced) and is coughed up from the lower respiratory tract i.e trachea and bronchi. Cytological examination of sputum is done under microscope to detect presence of malignant cells.
	
	\item{{\textbf{Abrasive Cytology}}}
	
	In this procedure, the sample is collected directly from the surface of the region of interest using superficial scraping or brushing of the lesion (artificial mechanical desquamation).
	
	\begin{enumerate}[]
		\item Scrape cytology: This technique deals with exfoliation of cells with the help of scrape or brush from the organ or the region being tested. Pap-smear test is a well known screening test of this kind. Buccal mucosal smear, skin scraping, esophagus, stomach, etc. also fall under this category.
		
		\item Brush cytology: It is used to collect cell samples from the gastrointestinal tract, bronchial tree, cervix etc.
		
	\end{enumerate}
\end{enumerate}
%\clearpage
{\textbf{Slide Preparation}
}

It is one of the important steps for the diagnosis of carcinoma from cytology images that is equally
necessary for both manual and automatic diagnosis system. After preparing the slide suitably, the
fixation is required. Two kinds of slide preparation techniques are normally performed in laboratory on the
collected specimen.
\begin{enumerate}[]
	\item{\textit{Conventional Preparations}}
	
	After collection, specimen is expelled into appropriately labelled glass slides with patient's unique
	identification. The expelled material is spread over several slides in small amounts, rather than
	deposited in one large pool on a single slide to enhance the probability of error free interpretation. This
	simplifies the process to obtain a thin-layer preparation. Spreading of the material over the slide is usually
	performed by another sliding glass slide, in order to avoid crushed artefacts and obtain a uniform smear.
	Large amounts of blood is avoided to prevent clotting and fibrin trapping in the cells,which creates large
	cracks on the slide hindering interpretation at cellular level. 
	\item{\textit{Liquid-Based Preparations}}
	
	It was introduced initially for cervical smears, and nowadays it is also used for other types of specimen,
	including FNA, because this technology has  added advantages compared to conventional
	smears. Three preparatory steps are undertaken viz.  cell dispersion, collection and transfer to prepare an appropriate slide. After
	collecting the specimen, the aspirate is rinsed directly into a container filled with 20 ml. of CytoLyt or
	CytoRich transport solution which is an alcohol-based solution (see Fig. \ref{fig:3}). If a fresh, non-alcohol-fixed specimen is indicated
	clinically, the specimen is put into a balanced electrolyte solution. Two commonly used liquid-based
	preparation techniques include ThinPrep (TP)\cite{dabbs1997immunocytochemistry} (Cytyc Corp, Marlborough, MA) and BD SurePath (SP)\cite{norimatsu2013cytologic}
	(TriPath Imaging Inc., Burlington, NC).
	
	Liquid based preparations  offers several  advantages over conventional preparations:\begin{itemize}
		\item Abundant cellularity in the specimen can be observed.
		\item Immediate liquid fixation avoids air-drying artifacts.
		\item Free from background contaminations like cell clumps, blood and mucus are very rare due to advanced preparatory techniques, give a good background clarity.
		\item Nuclear and cytoplasmic architecture are well maintained with reduced overlapping.
		\item Normalized specimen.
		\item It has potential capability for processing residual material as a cellblock.
	\end{itemize}
	
	\begin{figure}[h]
		\centering
		\includegraphics[width=0.5\linewidth]{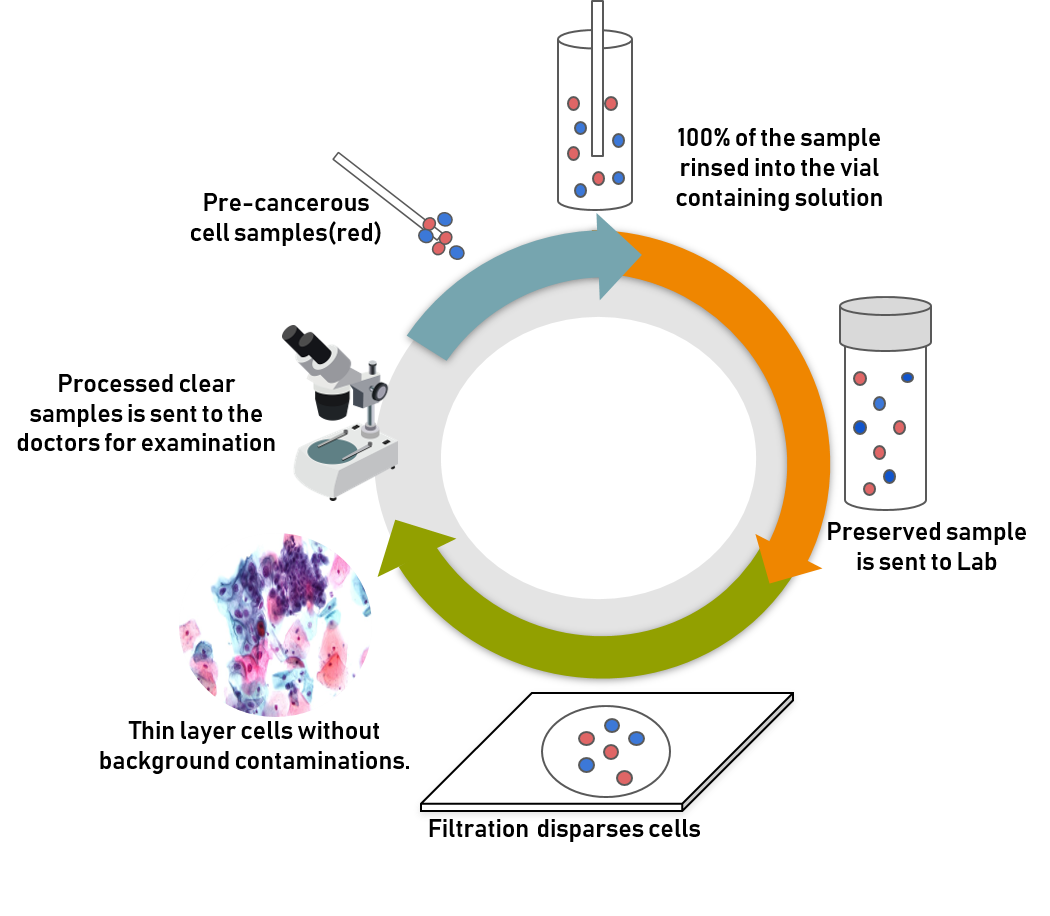}
		\caption{{Steps to prepare thin prep  cytology specimen }}
		\label{fig:3}
	\end{figure}
\end{enumerate}

\textbf{Fixation Techniques and Staining Protocol}

Immediate fixation of the collected specimen is crucial, otherwise it produces drying artifact leading to false positives or false negatives cases in medical diagnosis. Two types of  fixation  techniques are practiced in laboratories: Air drying and Alcohol drying.  Staining is also an essential step which is usually done after fixation of cytology specimen. Majority of the stains generally fascinates lights and illuminating samples under it. Without staining, it is not possible to identify the selective regions of different tissue samples. The quantity of illumination on a sample or portion of sample depends on the amount and type of stain. However, the same type of stain may vary on multiple factors such as manufacturer of stain, procedure of preservation and also the  condition of specimen before use, etc. \cite{5193250}. Even in {case of specimen expertise of} cytology, it may vary from one cytotechnologist to another. The intensity can vary depending upon the amount of time it remains under the air drying process.

\begin{enumerate}[]
	\item{\textit{Air Drying}}
	After preparation of the slides, they are immediately fixed by air drying
	preferably within 5 minutes. Romanowsky stains \cite{wittekind1983nature} are usually applied to air-dried smears. It's staining protocol includes May Grunwald Giemsa (MGG) \cite{cardozo1954clinical}, Leishman Giemsa(LG) \cite{doddagowda2017leishman}, Diff quick stain etc. MGG is the most commonly used staining technique to extract celluar morphology and cytoplasm details from air-dried smears.Tubercle bacilli, Actinomyces, some fungal elements  appear red and background  appear pale blue in color on application of this particular stain. 
	\item{\textit{Alcohol Fixation}}
	
	Alcohol or wet fixation is achieved either by using a spray fixative or dipping the slides in 95\% ethyl
	alcohol. Papanicolaou stain \cite{berkan1986protocol} is normally preferred for the staining of alcohol fixed slides. Squamous
	differentiation can be  best appreciated by Papanicolau stain. The staining protocal includes Papanicolaou-EA-50, to stain critical portions of nucleus and cytoplasm. Harris hematoxylin \cite{li2018hematoxylin} is a combination of  OG6 (Orange G) and EA50 (Eosin Azure). OG6 is a pap reagent for counter staining  exfoliative cytology samples like vaginal, cervical, prostatic smears etc. After application of the stain the nucleus  appears blue/black. The cytoplasm for keratinising cells  appears pink or orange and blue or green in colour for non-keratinising squamous cells.
	
	Some special stains  are mentioned in Table\ref{tab:tab}.
	
	\begin{table}[]
		\caption{{Some special stains and applicability}}
		\label{tab:tab}
		\centering
		
		\begin{tabular}{|l|l|}
			\hline
			\textbf{{Special Stains}}  & \textbf{{Requirement}}                                                           \\ \hline
			\begin{tabular}[c]{@{}l@{}} \small{{Modified Ziehl Neelson}}\\ \small{{\cite{allen1992modified}}}\end{tabular} & \begin{tabular}[c]{@{}l@{}}\small{{Acid fast bacilli smear and}} \\ \small{ {culture}}\end{tabular}
			\\ \hline
			
			\begin{tabular}[c]{@{}l@{}}\small{{Gram staining}}\\ \small{{\cite{gregersen1978rapid}}}\end{tabular} & \small{{Bacteria}}                                                                       \\ \hline
			\begin{tabular}[c]{@{}l@{}}\small{{Mucicarmine}}\\ \small{{\cite{cherry1938histopathological}}}\end{tabular}              & \small{{Mucins}}                                                                         \\ \hline
			\begin{tabular}[c]{@{}l@{}}\small{{PAS(Periodic Acid-Schif}} \\ \small{{\cite{warnock1988differentiation}}}\end{tabular} & \begin{tabular}[c]{@{}l@{}}\small{{Glycogen, Fungal wall,}}\\ \small{{Lipofuscin}}\end{tabular}   \\ \hline
			\begin{tabular}[c]{@{}l@{}}\small{{Oil red O} }\\\small{ { \cite{ramirez1992quantitation}}}\end{tabular}                & \small{{Lipids}}                                                                         \\ \hline
			\begin{tabular}[c]{@{}l@{}}\small{{Perl's Prussian blue}} \\ \small{{\cite{hall2013comparison}}}\end{tabular}     & \small{{Iron}}                                                                           \\ \hline
			\begin{tabular}[c]{@{}l@{}} \small{{Modified Fouchet's Test}} \\ \small{{\cite{bryant1955assessment}}}\end{tabular}   & \small{{Bilirubin}}                                                                     \\ \hline

		\end{tabular}
		
	\end{table}
	
\end{enumerate}

\bibliography{ref}
\bibliographystyle{plain}

\end{document}